\documentclass[journal=nalefd,manuscript=letter]{achemso}
\usepackage[version=4]{mhchem}
\usepackage{amsmath}
\usepackage{booktabs}
\usepackage{graphicx}

\graphicspath{{figures/}}

\makeatletter

\providecommand*{\@mcitecorrectmaxwidthbibitem}{0}
\providecommand*{\@mcitecorrectmaxwidthsubitem}{0}
\makeatother

\usepackage[mathlines]{lineno}
\mciteErrorOnUnknownfalse

\title{Nonadiabatic molecular dynamics simulation with external field}
\author{Chengji Li}

\author{Fan Zheng}
\affiliation{School of Physical Science and Technology, ShanghaiTech University, Shanghai 201210, China\relax}
\email{zhengfan@shanghaitech.edu.cn}

\begin{document}

\begin{abstract}
Nonadiabatic molecular dynamics simulations with classical path approximation (CPA-NAMD) cannot simulate the carrier dynamics under an external field.
We develop CPA-NAMD-extF, a first-principles density-matrix framework for calculating field-driven electronic responses along precomputed finite-temperature ionic trajectories.
The method combines explicit external-field coupling with self-consistent updates of electronic density while retaining the computational efficiency of the classical-path approximation.
Benchmarks against regular TDDFT methods show close agreement for time-dependent dipoles, finite-field currents, and linear optical responses.
Applications to the linear absorption spectrum of crystalline and amorphous SiO$_2$ demonstrate its capability to large and structurally complex systems.
This method is also applied to the 2D hybrid perovskite $(\mathrm{BA})_2\mathrm{CsPb}_2\mathrm{Br}_7$.
Finite-temperature calculations show that the second-order opto-current response is strongly affected by structural fluctuations.
Real-space DC current density further reveals pronounced Br-centered vortex currents, and its microscopic origin is investigated.
\end{abstract}

\section*{I. Introduction}
Nonadiabatic molecular dynamics (NAMD) has been widely used to investigate various excited carrier dynamics in extended systems,~\cite{akimov_pyxaid_2013,zheng_multiple_2023} 
such as carrier cooling~\cite{zheng_multiple_2023,zhou_control_2022,wei_long-lived_2023,boehme_phonon-mediated_2020,yin_tuning_2019,hu_reducing_2024,zhou_enhancing_2024}, interfacial charge transfer~\cite{zheng_phonon-assisted_2017,zheng_ultrafast_2019,chu_ultrafast_2022,tao_tunable_2024,zhang_ultrafast_2024,zheng_carrier_2024}, photoinduced charge separation~\cite{shi_rapid_2023,zhang_nuclear_2024,liao_strain_2025,yan_dramatically_2025,wang_rapid_2021}, and carrier transport~\cite{zheng_carrier_2024,yang_interlayer_2024,guo_strong_2021}.
In particular, by employing the classical path approximation (CPA), the electronic dynamics simulated by NAMD (CPA-NAMD) are propagated as a post-processing step along a precomputed ground-state Born--Oppenheimer molecular-dynamics (BOMD) trajectory, while the feedback of electronic excitation on the nuclear motion is neglected.
Consequently, the major computational cost arises from the ground-state BOMD simulation, whereas the subsequent CPA-NAMD propagation requires only a relatively small additional cost while retaining the first-principles DFT description.~\cite{kang_nonadiabatic_2019,zheng_ultrafast_2019}
However, the current CPA-NAMD implementation does not explicitly describe carrier excitation driven by an external field.
This is because CPA-NAMD uses Hamiltonians obtained from precomputed ground-state calculations and does not explicitly include the external-field induced modification of the electronic potential.
Therefore, this limitation restricts its applicability to field-induced electronic and optical responses.

To describe electronic excitation and current responses driven by external fields in real time, real-time time-dependent density-functional theory (rt-TDDFT) usually provides a direct first-principles approach. 
In periodic systems, the time-dependent current obtained from rt-TDDFT can be used to extract linear optical responses and nonlinear photocurrents.~\cite{bertsch_real-space_2000,yabana_time-dependent_2012,noda_salmon_2019,he_ultrafast_2024}
However, most real-space or plane-wave rt-TDDFT implementations typically require a very small electronic time step for accurate and stable propagation of the time-dependent Kohn--Sham (KS) orbitals.
The resulting computational cost becomes substantial for systems containing hundreds of atoms and for long propagation times.~\cite{jia_fast_2018}
Finite-temperature response calculations additionally need electronic propagations over many thermally sampled trajectory segments in order to obtain statistically reliable results.~\cite{jacobs_impact_2024,jacobs_pulse-induced_2023}
Separating different nonlinear orders in finite-field calculations can also require reversed fields or multiple field amplitudes.~\cite{kim_giant_2022,ofstad_adiabatic_2023}

In this work, we introduce CPA-NAMD-extF, an advancement of CPA-NAMD which can explicitly incorporate external-field coupling into the electronic dynamics. 
By implementing CPA and propagating the electronic density matrix using the ground-state BOMD trajectories, CPA-NAMD-extF substantially reduces the cost of calculations compared to regular rt-TDDFT simulations. 
This method allows us to efficiently simulate the external-field-induced optical current for systems containing more than 500 atoms.
We benchmark the time-dependent dipoles of small molecules, finite-field currents, and linear optical responses against existing rt-TDDFT implementations to validate our implementation. 
Then, we use this method to compute the absorption spectra of crystalline and amorphous SiO$_2$ systems, which are usually challenging for rt-TDDFT calculations.
This method is further applied to investigate the finite-temperature shift-current response of the 2D hybrid perovskite $(\mathrm{BA})_2\mathrm{CsPb}_2\mathrm{Br}_7$.
Moreover, the vortex-type shift current in real space is also investigated to reveal its microscopic origin.

\section*{II. Method and Implementation}
CPA-NAMD treats the nuclear motion classically while propagating the electronic dynamics quantum mechanically along a precomputed ground-state BOMD trajectory.~\cite{kang_nonadiabatic_2019,zheng_ultrafast_2019} 
At each BOMD step, the instantaneous adiabatic Kohn--Sham states and eigenenergies are obtained from ground-state DFT calculations and they are used to construct the electronic Hamiltonian. 
The electronic wavefunctions or density matrix is then propagated along the prescribed Hamiltonian. 
With CPA, the feedback of the excited carriers on the nuclear motion is neglected, allowing the electronic dynamics to be evaluated as a post-processing step of the ground-state BOMD simulation.
CPA is widely used in NAMD simulations, particularly in periodic bulk or low-dimensional materials.
This is because these materials usually have heavy ions whose motions are much slower than the process of carrier excitation and transport.
Meanwhile, the time-evolving wavefunctions are also mostly delocalized across the material.
Such additional force by the wavefunction is relatively weak.
In our implementation of NAMD with an external field, CPA is also used.
Atomic units are used throughout this paper unless otherwise stated.

In CPA-NAMD-extF, the ions evolve along a precomputed BOMD trajectory, $\mathbf R(t)$, while the electronic density matrix is explicitly propagated under an external field along such trajectory. 
During the simulation, the propagated electronic density and the associated Hartree and exchange--correlation potentials are updated self-consistently at each electronic step. 
By assuming CPA, for the relatively weak external fields considered here, the excitation-induced ionic displacements are small compared with the underlying thermal fluctuations.
For each BOMD configuration, we extract the ground-state KS Hamiltonian $H_0$, the ground-state Hartree--exchange-correlation potential matrix $V_{0,\mathrm{Hxc}}$, and the velocity matrix $\mathbf v$, together with the overlap matrices between KS states at neighboring MD configurations. 
Here, $H_0$, $V_{0,\mathrm{Hxc}}$, and $\mathbf v$ are computed with the BOMD adiabatic states as the basis.
Thus,
\begin{equation}
H_{0,ij}^{(t_m)}(t_m)
=
\varepsilon_i(t_m)\delta_{ij},
\end{equation}
where $t_m$ denotes the $m$-th BOMD step, the superscript $t_m$ indicates the adiabatic basis used the operator, and $\varepsilon_i(t_m)$ is the eigenenergy obtained at $t_m$ during the BOMD calculation. 
Similarly, the ground-state Hamiltonian at the next BOMD step $t_{m+1}$ represented in the adiabatic basis at $t_m$ is
\begin{equation}
H_0^{(t_m)}(t_{m+1})
=
U(t_{m+1}, t_m)
H_0^{(t_{m+1})}(t_{m+1})
U(t_{m+1}, t_m)^{\dagger},
\end{equation}
where $U_{ij}(t_{m+1}, t_m)=\langle \phi_i(t_m)|\phi_j(t_{m+1}) \rangle$ is the overlap between the adiabatic states of the two neighboring BOMD steps. 
The ground-state $V_{0,\mathrm{Hxc}}$ and the velocity matrix $\mathbf v$ are treated in the same manner. 
These quantities are calculated within BOMD with negligible additional computational costs. 
Following the same idea as conventional CPA-NAMD,~\cite{zheng_ultrafast_2019,zheng_carrier_2024,ren_nonadiabatic_2013} the ground-state $H_0$, $V_{0,\mathrm{Hxc}}$, and $\mathbf v$ are linearly interpolated within each BOMD step to obtain $H_0(t)$, $V_{0,\mathrm{Hxc}}(t)$, and $\mathbf v(t)$ on the finer electronic time grid (SI, Eq.~\ref{si-eq:matrix-interpolation}).
Because the ionic configuration changes only slightly over each short BOMD interval, the corresponding ground-state charge density and the electronic matrices vary smoothly and can be locally approximated by linear interpolation.\cite{ren_nonadiabatic_2013}

During the electronic propagation, the electronic excitations alter the Hartree--exchange--correlation potential from its ground-state value. 
Therefore, we replace the ground-state Hartree--exchange--correlation contribution contained in the KS Hamiltonian with the time-dependent potential constructed from the propagated electronic density.
The propagation Hamiltonian is thus written as
\begin{equation}
H_{\mathrm{prop}}(t)
=
H_0(t)
-
V_{0,\mathrm{Hxc}}(t)
+
V_{\mathrm{Hxc}}[\rho(t)]
+
V_{\mathrm{field}}(t).
\label{eq:hprop-main}
\end{equation}
Here $V_{\mathrm{Hxc}}[\rho(t)]$ is updated self-consistently from the propagated density $\rho(t)$ at each electronic time step. 
In our implementation, the external field is introduced in the velocity gauge through a time-dependent vector potential $\mathbf A(t)$, related to the electric field by $\mathbf E(t)=-c^{-1}\partial\mathbf A(t)/\partial t$, where $c$ is the speed of light.~\cite{yabana_time-dependent_2012,pemmaraju_velocity-gauge_2018}
The field-coupling potential is
\begin{equation}
V_{\mathrm{field}}(t)
=
\frac{1}{c}
\mathbf A(t)\cdot\mathbf v(t).
\label{eq:field-main}
\end{equation}
Here, $\mathbf v(t)$ denotes the velocity matrix evaluated in the adiabatic KS basis. 
The corresponding velocity operator, defined as
$\hat{\mathbf v}
=
\hat{\mathbf p} -i \left[ \hat{\mathbf r},\hat V_{\mathrm{ion}}^{\mathrm{nl}}
\right]$,
includes the nonlocal part of the ionic pseudopotential $(\hat V_{\mathrm{ion}}^{\mathrm{nl}})$ and $\hat{\mathbf p}$ is the momentum operator. 

Within each BOMD step from $t_m$ to $t_{m+1}$, the adiabatic basis $\phi(t_m)$ is used throughout this step.
We split one BOMD step into several electronic steps.
The density matrix and the corresponding charge density are evolved self-consistently within each electronic step using a procedure similar to that reported in Ref.\kern-0.42em\citenum{wang_efficient_2015}.
Specifically, at each electronic step, a trial charge density is extrapolated from the previous steps, and it is used to construct $V_{\mathrm{Hxc}}[\rho]$ and the Hamiltonian including external fields at the end of electronic step. 
With both Hamiltonians constructed for the start and end of this electronic step, the propagating Hamiltonian is further linearly interpolated within this electronic step.
The density matrix is then propagated under this Hamiltonian until the end of electronic step; 
A new electronic charge density is reconstructed from the propagated density matrix and the adiabatic basis. 
The evolved charge density is mixed with the trial charge density to generate a new charge density for the end of the electronic step.
This procedure is repeated until a self-consistent charge density is obtained.
In our implementation, the time step of electronic step is around 0.1fs to ensure a stable and efficient calculation. 
With the external field, self-consistency of the propagated charge density is achieved within 4 to 5 iterations.

To reduce the computational cost, not all the valence and conduction adiabatic states are used for evolution of charge densities. 
Instead, we truncate the adiabatic-state space and divide it into active $(\mathcal A)$ and inactive $(\mathcal I)$ subspaces. 
Under the relatively weak fields considered here, appreciable population changes are expected mainly within the excitation window (active subspace), whereas the deep-valence states (inactive subspace) remain nearly fully occupied. 
Their occupations are therefore fixed to the corresponding BOMD adiabatic-state occupations.
In our implementation, the density matrix in the active subspace is propagated explicitly as
\begin{equation}
i\frac{dD_{\mathcal A\mathcal A}(t)}{dt}
=
\left[
H_{\mathrm{prop}}(t),
D_{\mathcal A\mathcal A}(t)
\right].
\label{eq:density-matrix-main}
\end{equation}
Here, a much finer time step around 100as is used for the propagation. 
Since both $H_{\mathrm{prop}}(t)$ and $D_{\mathcal A\mathcal A}(t)$ are expressed using the adiabatic states, the time cost of such propagation is negligible.
The charge density entering $V_{\mathrm{Hxc}}[\rho(t)]$ is constructed from the propagated active-space density matrix $D_{\mathcal A\mathcal A}(t)$ together with the contribution of the inactive states evaluated using their fixed BOMD adiabatic-state occupations.
The inactive density matrix $D_{\mathcal I\mathcal I}(t)$, composed of deep-valence states, is kept fixed as a diagonal matrix without propagation.
However, the field-induced coherence $D_{\mathcal I\mathcal A}(t)$ cannot be kept fixed.
To account for coupling between the inactive and active subspaces, it is evaluated at each electronic time step using an instantaneous first-order adiabatic approximation (see Sec.~\ref{si-sec:si-method} of the Supplemental Information for more details).
We find that such coherence contributes to the final field-induced current, but it is not necessary to be included in the self-consistent density or potential evolution.

With the electronic density matrix constructed above, the current induced by the external field can be evaluated.
The macroscopic cell-averaged current density is written as $\mathbf J(t)=1/\Omega \, \mathbf j(t)$, and 
\begin{equation}
\mathbf j(t)
=
\mathbf j^{\mathrm{para}}(t)
+
\mathbf j^{\mathrm{dia}}(t)
+
\mathbf j^{\mathrm{corr}}(t),
\label{eq:current-main}
\end{equation}
where $\Omega$ is the unit-cell volume.
The paramagnetic contribution is evaluated from the electronic density matrix and the velocity matrix as
\begin{equation}
\mathbf j^{\mathrm{para}}(t)
=
-
\operatorname{Re}\{
\operatorname{Tr}
\left[
D(t)\mathbf v(t)
\right]
\}.
\label{eq:paramagnetic-current}
\end{equation}
where the trace denotes a sum over the adiabatic states.
Here, the density matrix $D(t)$ is assembled in the full active-inactive block as
\begin{equation}
D(t)
=
\begin{pmatrix}
D_{\mathcal I\mathcal I} & D_{\mathcal I\mathcal A}(t)\\
D_{\mathcal A\mathcal I}(t) & D_{\mathcal A\mathcal A}(t)
\end{pmatrix}.
\label{eq:density-matrix-block}
\end{equation}
The diamagnetic contribution is $\mathbf j^{\mathrm{dia}}(t)=-\frac{N_e}{c}\mathbf A(t)$, where $N_e$ is the number of electrons in the unit cell.
In the velocity gauge, convergence of the current generally requires a large number of high-energy unoccupied states.
Since only a finite number of bands are included in the present calculation, we employ a first-order adiabatic correction derived in Ref.\kern-0.42em\citenum{yakovlev_adiabatic_2017} to compensate for the missing high-energy-band contributions caused by the finite-band truncation.
The correction is given by
\begin{equation}
\mathbf j^{\mathrm{corr}}(t)
=
\frac{1}{c}
\sum_{n\in\mathrm{occ}}
\left[
\mathbf A(t)
-
2
\sum_{\substack{i=1\\i\neq n}}^{N_{\mathrm{b}}}
\frac{
\operatorname{Re}
\left[
\left(
\mathbf v_{ni}(t)\cdot\mathbf A(t)
\right)
\mathbf v_{in}(t)
\right]
}{
\varepsilon_i(t_m)-\varepsilon_n(t_m)
}
\right].
\label{eq:finite-band-correction}
\end{equation}
Here, $N_{\mathrm b}$ is the number of bands included in the current evaluation, and the sum over $n$ runs over the occupied reference states.
The correction is evaluated from the adiabatic energies and velocity matrix and is linear in the vector potential.
Further details of the evaluation of $D_{\mathcal I\mathcal A}(t)$ and the finite-band correction are provided in Sec.~\ref{si-sec:si-method} of the Supplemental Information.

To calculate the linear optical absorption spectrum, we apply a weak $\delta$-kick perturbation in the velocity gauge.
The imaginary part of the dielectric function, $\varepsilon_2(\hbar\omega)$, is obtained from the Fourier-transformed current and normalized using an effective $f$-sum rule.
The real part, $\varepsilon_1(\hbar\omega)$, is then obtained through the Kramers--Kronig relation.\cite{kuzmenko_kramerskronig_2005}
The absorption coefficient spectrum is thus calculated as
\begin{equation}
\alpha(E_{\mathrm{ph}})
=
\frac{4\pi E_{\mathrm{ph}}}{hc}
\sqrt{
\frac{
\sqrt{\varepsilon_1^2(E_{\mathrm{ph}})+\varepsilon_2^2(E_{\mathrm{ph}})}
-
\varepsilon_1(E_{\mathrm{ph}})
}{2}
},
\label{eq:absorption-main}
\end{equation}
where $E_{\mathrm{ph}}=\hbar\omega$ is the photon energy.
Further details of the spectral analysis are provided in Sec.~\ref{si-sec:si-response} of the Supplemental Information.

Moreover, the field-induced currents allow us to extract higher-order nonlinear responses to the external field, such as the shift current response.  
The field-induced even-order current can be isolated from the positive-field, negative-field, and zero-field calculations as
\begin{equation}
\mathbf{J}_{\mathrm{even}}(t)
=
\frac{
\mathbf{J}(t;+E_0)
+
\mathbf{J}(t;-E_0)
}{2}
-
\mathbf{J}(t;0).
\label{eq:even-order-current}
\end{equation}
Here, $E_0$ is the amplitude of the applied electric field. The $\mathbf{J}(t;0)$ term removes the field-independent background, while the $+E_0/-E_0$ combination cancels all odd-order contributions. 
In the weak-field regime,
\begin{equation}
\mathbf{J}_{\mathrm{even}}(t)
=
E_0^2 \mathbf{J}^{[2]}(t)
+
\mathcal O(E_0^4),
\label{eq:even-order-expansion}
\end{equation}
and the extracted signal is dominated by the second-order response, i.e. the shift current response.
For a monochromatic field at frequency $\omega_0$, the direct-current (DC) second-order response corresponds to the zero-frequency component of the current. 
This component is obtained by averaging the computed even-order current over a time window as $\mathbf{J}_{\mathrm{DC}}=(t_b-t_a)^{-1}\int_{t_a}^{t_b}\mathbf{J}_{\mathrm{even}}(t)\,dt$.
With the DC current computed, the corresponding second-order conductivity is then evaluated as
\begin{equation}
\sigma^{(2)}(0;\omega_0,-\omega_0)
=
\frac{\mathbf{J}_{\mathrm{DC}}}{E_0^2}.
\label{eq:second-order-conductivity}
\end{equation}
Here, $t_a$ and $t_b$ denote the beginning and end of the averaging window, respectively. 
The window is chosen within the constant current-envelope region to span an integer number of external-field periods, i.e., $t_b-t_a=N(2\pi/\omega_0)$, where $N$ is an integer.

Since the BOMD adiabatic states are used as the basis in our simulations, it is possible to obtain the real-space current density $\mathbf j(\mathbf r,t)$.~\cite{hu_light-induced_2023}
The position-resolved current density operator is expressed as $\hat{\mathbf j}(\mathbf r)=-1/2\left[|\mathbf r\rangle\langle\mathbf r|\hat{\mathbf v}+\hat{\mathbf v}^{\dagger}|\mathbf r\rangle\langle\mathbf r|\right]$.
Thus, we can construct the real-space current density on the FFT grid used in the DFT calculation.
Combining the paramagnetic, diamagnetic, and finite-band correction contributions, it is evaluated as
\begin{equation}
\mathbf{j}(\mathbf r,t)
=
\operatorname{Tr}
\left[
\left(
D(t)
+
\delta D_c(t)
\right)
\mathbf{j}(\mathbf r)
\right]
+
\frac{\mathbf A(t)}{c}
\left[
\rho_0(\mathbf r;t)
-
\rho(\mathbf r,t)
\right].
\label{eq:real-space-current-trace}
\end{equation}
Here, $\hat{\mathbf j}(\mathbf r)$ is position-resolved current operator, and the trace is taken over the adiabatic states of the combined active--inactive space $\mathcal R=\mathcal I\oplus\mathcal A$.
$\delta D_c(t)$ denotes the finite-band adiabatic correction (SI, Eq.~\ref{si-eq:realspace-delta-density-correction}).
The reference density $\rho_0(\mathbf r;t)$ is the ground-state charge density at the same ionic configuration, whereas $\rho(\mathbf r,t)$ is the corresponding time-dependent electronic density.
The second term combines the diamagnetic contribution with the occupation-sum part (i.e., first term) of the finite-band correction (SI, Eq.~\ref{si-eq:current-correction}).
A detailed derivation is provided in Sec.~\ref{si-sec:si-real-space-current} of the Supplemental Information.
By construction, the resulting real-space current density satisfies
\begin{equation}
\mathbf{J}(t)
=
\frac{1}{\Omega}
\int_\Omega
\mathbf{j}(\mathbf r,t)\,d^3r.
\label{eq:local-macro-current}
\end{equation}
For the even-order real-space current density, the density matrix in Eq.~\eqref{eq:real-space-current-trace} is replaced by $D_{\mathrm{even}}(t)$, where $D_{\mathrm{even}}(t)=[D(t;+E_0)+D(t;-E_0)]/2-D(t;0)$.
Thanks to the CPA, $D_{\mathrm{even}}(t)$ can be obtained without additional basis transformations.
This is because $D(t;+E_0)$, $D(t;-E_0)$, and $D(t;0)$ are computed in the same BOMD adiabatic basis.

Our CPA-NAMD-extF method is implemented based on the pseudopotential plane-wave Quantum ESPRESSO (QE) package.~\cite{giannozzi_quantum_2009,giannozzi_advanced_2017}
We modify QE to export the adiabatic KS wave functions and eigenenergies, the ground-state Hartree--exchange--correlation potential and velocity matrices, and the overlap matrices between adiabatic states of neighboring BOMD configurations.
CPA-NAMD-extF reads these quantities and interfaces with QE routines to reconstruct the charge density and update the Hartree--exchange--correlation potential on the FFT grid. 
The implementation supports hybrid MPI/OpenMP parallelization.
The stored BOMD data can be reused for external fields with different amplitudes, polarizations, and temporal profiles.

Before we proceed, we first benchmark our CPA-NAMD-extF implementation against existing rt-TDDFT codes to demonstrate its validity. 
For the time-dependent dipole response, CPA-NAMD-extF closely reproduces the PWmat rt-TDDFT~\cite{jia_analysis_2013,jia_fast_2013,wang_efficient_2015} results for H$_2$CO$_3$ single molecule, 1D C$_{14}$H$_{12}$BN conjugated chain, and 2D monolayer WS$_2$, as shown in Fig.~\ref{si-fig:dipole-benchmark}. 
These tests show that the charge density evolution under the external field in our implementation is consistent with existing rt-TDDFT methods.
For monolayer MoS$_2$, the finite-field time-dependent current obtained with CPA-NAMD-extF can closely reproduce the SALMON reference result in Fig.~\ref{si-fig:mos2-current-benchmark}.~\cite{noda_salmon_2019} 
With the external field induced current calculation, we further compute the absorption spectrum bulk for crystalline Si, i.e. the linear optical response extracted from the current induced by a weak $\delta$-kick type external potential. 
Our result agrees well with the PWmat calculation as shown in Fig.~\ref{si-fig:si-linear-benchmark}.

\section*{III. Results}
We next apply CPA-NAMD-extF to the finite-temperature linear optical response of large and complex systems. 
Specifically, we calculate the absorption spectra of crystalline and amorphous bulk SiO$_2$ at 300 K. 
Each supercell contains 576 atoms and 3072 valence electrons. 
The crystalline model is constructed from an $\alpha$-quartz-derived supercell, whereas the amorphous model is generated using a Monte Carlo bond-switching procedure.~\cite{zheng_effects_2017} 
The BOMD trajectories are generated using QE with the Perdew--Burke--Ernzerhof (PBE) generalized-gradient approximation~\cite{perdew_generalized_1996}, SG15 norm-conserving pseudopotentials~\cite{schlipf_optimization_2015}, and $\Gamma$-point Brillouin-zone sampling.
For the real-time electronic propagation, the active spaces span bands 961--1950 for crystalline SiO$_2$ and 906--1990 for amorphous SiO$_2$.  
For each system, a weak $\delta$-kick external potential of $\Delta k=0.002~\mathrm{bohr}^{-1}$ is applied independently along the $x$, $y$, and $z$ directions.
Here, the final absorption spectrum is obtained by averaging the three polarization-resolved spectra.
\begin{figure}[!htbp]
\centering
\includegraphics[width=\linewidth]{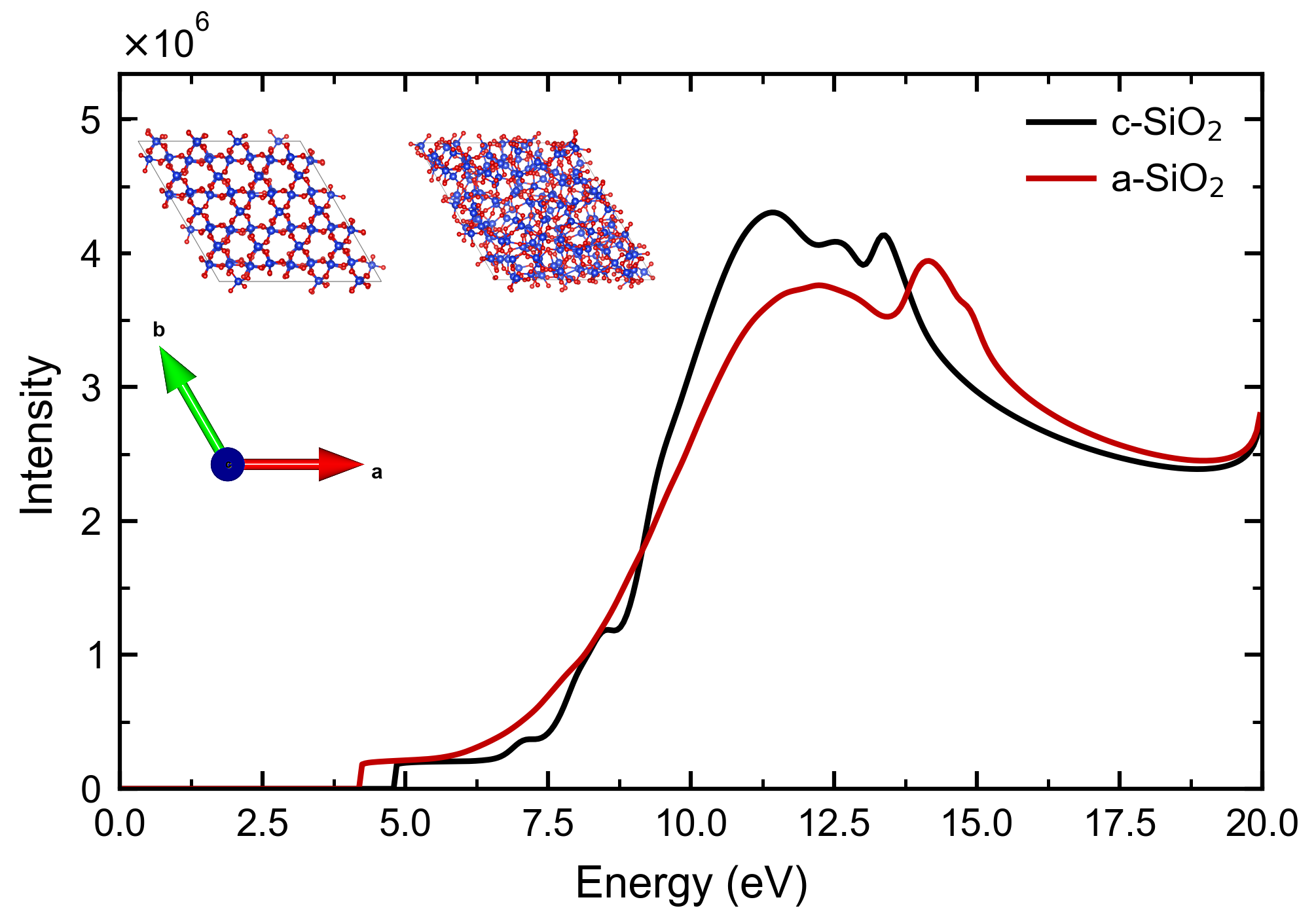}
\caption{
Absorption spectra of crystalline and amorphous 576-atom SiO$_2$ models at 300 K calculated with CPA-NAMD-extF. 
The spectra are obtained by averaging the independently calculated $x$-, $y$-, and $z$-polarized weak $\delta$-kick responses. Absorption intensities are shown in arbitrary units.
}
\label{fig:sio2-absorption}
\end{figure}
As shown in Fig.~\ref{fig:sio2-absorption}, the amorphous structure exhibits broader and smoother absorption features than the crystalline one, together with a more pronounced low-energy absorption tail.
This is consistent with previous experimental studies of crystalline and amorphous SiO$_2$.~\cite{tan_optical_2005,tamura_first-principles_2008,sadigh_first-principles_2011}
More importantly, the present calculation demonstrates that CPA-NAMD-extF can treat current-based optical responses in structurally ordered or disordered cells containing hundreds of atoms along explicit finite-temperature ionic trajectories.
Such systems are computationally demanding, requiring several days of computation on multiple GPUs using conventional rt-TDDFT, particularly for finite-temperature trajectory sampling. 
In comparison, the present CPA-NAMD-extF calculations for the 576-atom crystalline and amorphous SiO$_2$ systems only require hours on CPU cores.

\begin{figure}[!htbp]
\centering
\includegraphics[width=0.95\linewidth]{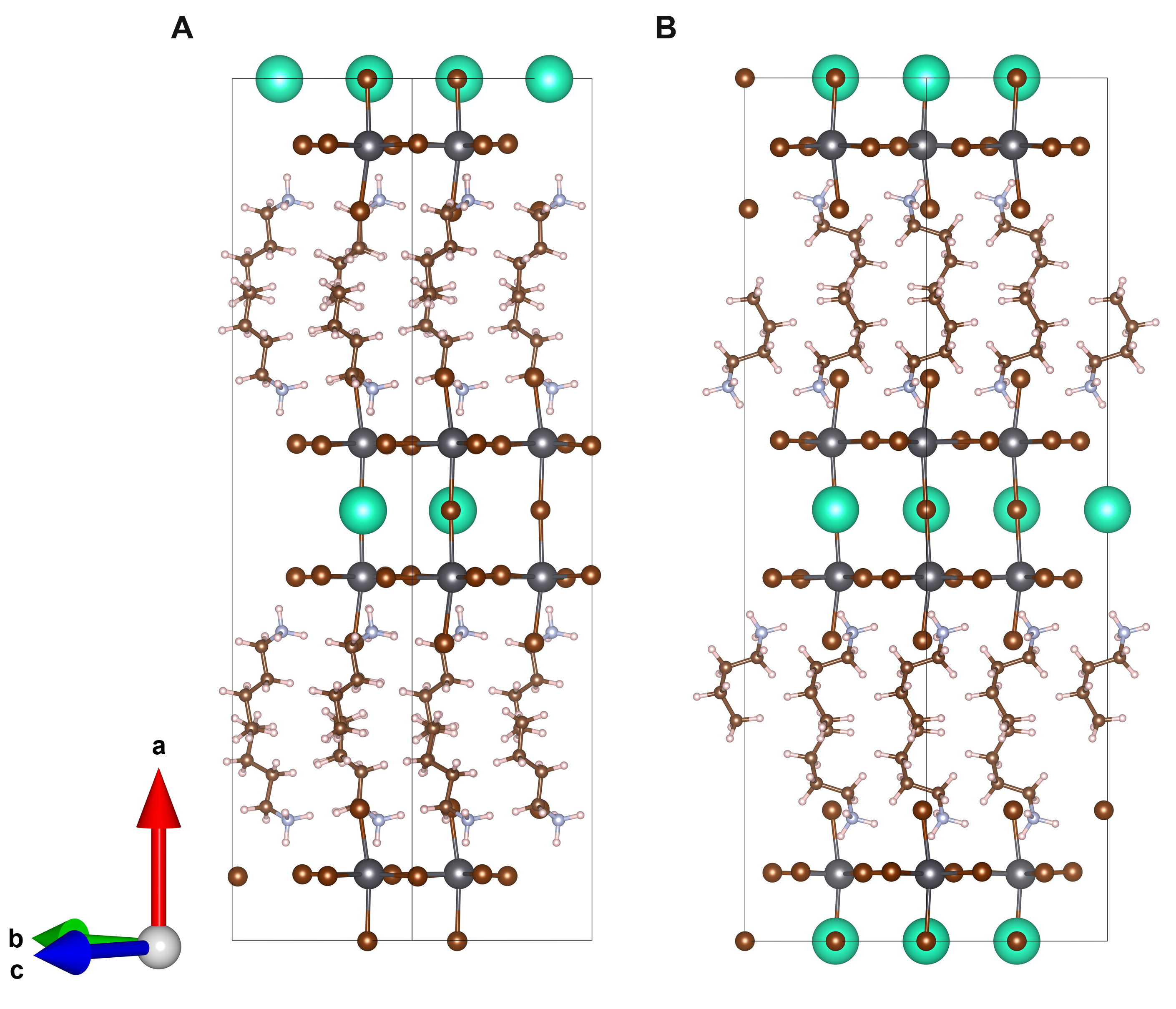}
\caption{
Relaxed structure models of $(\mathrm{BA})_2\mathrm{CsPb}_2\mathrm{Br}_7$.
(A) Ferroelectric structure used to initialize the 300 K simulation.
(B) Paraelectric structure used to initialize the 450 K simulation.
Both models are $2\times2\times2$ supercells.
For both structures, the Cartesian $z$ direction used for the field-response calculations is aligned with the crystallographic $\vec{b} + \vec{c}$ direction.
}
\label{fig:bacspbbr-structures}
\end{figure}

As the main nonlinear-response application, we investigate the 2D hybrid perovskite $(\mathrm{BA})_2\mathrm{CsPb}_2\mathrm{Br}_7$, where $\mathrm{BA}=\mathrm{C_4H_9NH_3^+}$.
This material is experimentally reported to undergo a ferroelectric-to-paraelectric transition at the Curie temperature $T_c=412$ K.~\cite{wu_alloying_2018}
Following the experimentally determined ferroelectric and paraelectric structures reported in Ref.\kern-0.42em\citenum{wu_alloying_2018}, we construct a $2\times2\times2$ supercell for each phase.
Each structure contains 352 atoms and 1216 valence electrons. The ferroelectric and paraelectric structures are shown in Fig.~\ref{fig:bacspbbr-structures}(A) and (B), respectively.
In the ferroelectric structure, the ordered arrangement of the BA$^+$ cations induces an asymmetric distortion of the inorganic framework through ordered hydrogen-bonding interactions, establishing a polar axis along the crystallographic $\vec{b} + \vec{c}$ direction.
By contrast, the paraelectric structure reflects orientational disorder of the BA$^+$ cations and a higher average symmetry of the inorganic framework.
For the subsequent field-response calculations, the Cartesian $z$ axis is chosen along the crystallographic $\vec{b} + \vec{c}$ direction for both the ferroelectric and paraelectric phases.
BOMD calculations are performed using QE with the PBE functional, SG15 norm-conserving pseudopotentials~\cite{schlipf_optimization_2015}, DFT-D3 dispersion correction~\cite{grimme_consistent_2010}, and $\Gamma$-point Brillouin-zone sampling.
The 300 and 450 K simulations are initialized from the ferroelectric and paraelectric structures, respectively.
For the real-time electronic propagation, the active spaces span bands 193--1187 of the ferroelectric structure and 193--1243 of the paraelectric structure.

Because 300 and 450 K lie below and above the experimentally reported Curie temperature, respectively, the two trajectories provide finite-temperature simulations on opposite sides of the phase transition.
For each temperature, we first equilibrate the system for 4 ps in the NVT ensemble using ab initio MD, followed by a 4 ps NVE production trajectory used for response sampling.
Owing to the high computational expense of the nonlinear response calculations, we do not perform the current calculation over the full 4 ps trajectory.
Instead, 51 snapshots are sampled at an interval of 80 fs.
This is because the nonlinear response is determined by the atomic structure, which evolves on picosecond and subpicosecond time scales.
An 80fs interval can still show a smooth evolution of shift current response.
Starting from each snapshot, a 20 fs field-driven electronic propagation is performed along the corresponding ionic trajectory segment.
Across the 51 sampled configurations, the average PBE HOMO--LUMO gaps are 2.56 eV at 300 K and 2.38 eV at 450 K.
The electronic density matrix is propagated with a time step of $0.1~\mathrm{fs}$.
Positive-field, negative-field, and zero-field calculations are carried out along the Cartesian $z$ direction using an optical field with a photon energy of 4.0 eV and a peak field amplitude of $0.01~\mathrm{V}/\text{\AA}$ (SI, Eq.~\ref{si-eq:monochromatic-vector-potential}).
The three time-dependent current traces are combined to isolate the even-order current response.
Its quadratic scaling with field amplitude is verified in the weak-field regime, as shown in the SI (Fig.~\ref{si-fig:weak-field-convergence}), confirming that the response is dominated by the second-order contribution.
The second-order conductivity $\sigma^{(2)}_{zzz}$ is then extracted from this response.

\begin{figure}[!t]
  \centering
  \includegraphics[width=1.0\linewidth]{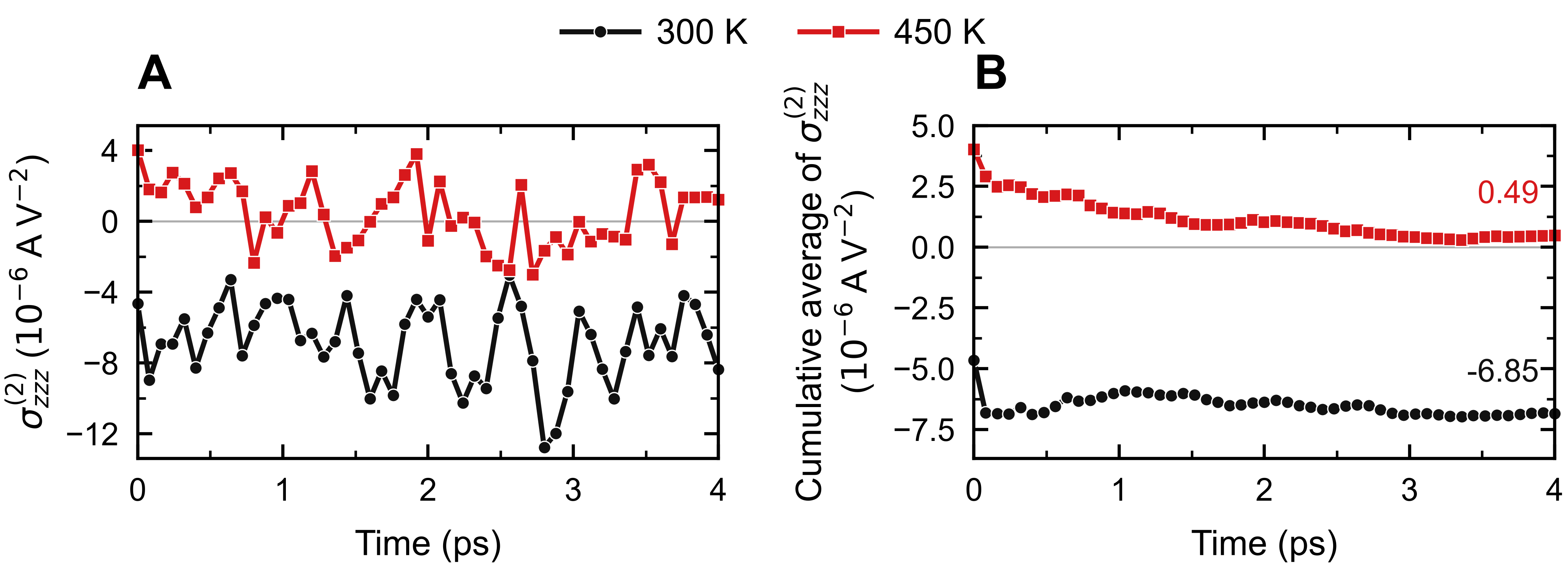}
  \caption{
  Finite-temperature second-order conductivity of
  $(\mathrm{BA})_2\mathrm{CsPb}_2\mathrm{Br}_7$ calculated with
  CPA-NAMD-extF.
  (A) $\sigma^{(2)}_{zzz}(t)$ along the 300 and 450 K
  NVE trajectories.
  (B) Corresponding cumulative trajectory averages.
  Black circles and red squares denote the 300 and 450 K results,
  respectively.
  }
  \label{fig:bacspbbr-response}
\end{figure}

The finite-temperature second-order response is shown in Fig.~\ref{fig:bacspbbr-response}. 
Figure~\ref{fig:bacspbbr-response}A presents the $\sigma^{(2)}_{zzz}$ values obtained along the 300K and 450 K NVE trajectories. 
At 300 K, the response remains negative throughout the trajectory and fluctuates around a clearly nonzero mean, indicating that the sampled thermal configurations generally exhibit a nonvanishing nonlinear response. 
By contrast, the response at 450 K has an overall smaller magnitude and frequently changes sign. 
Although individual thermal configurations along the high-temperature trajectory can still generate finite second-order responses, these responses tend to vanish when averaged over a relatively long time range.
The different statistical behaviors of the two trajectories are more clearly revealed by the cumulative averages in Fig.~\ref{fig:bacspbbr-response}B. 
At 300 K, the cumulative average remains close to a finite value over most of the sampled trajectory and reaches $-6.85\times10^{-6}~\mathrm{A/V^2}$ at 4 ps.
By contrast, the cumulative average at 450 K gradually approaches zero from an initially positive value, indicating progressive cancellation between positive and negative contributions. 
These results demonstrate that a single static configuration cannot reliably represent the macroscopic finite-temperature nonlinear response, highlighting the need to explicitly sample and average over multiple thermal configurations along finite-temperature trajectories.

\begin{figure}[!htbp]
\centering
\includegraphics[width=\linewidth]{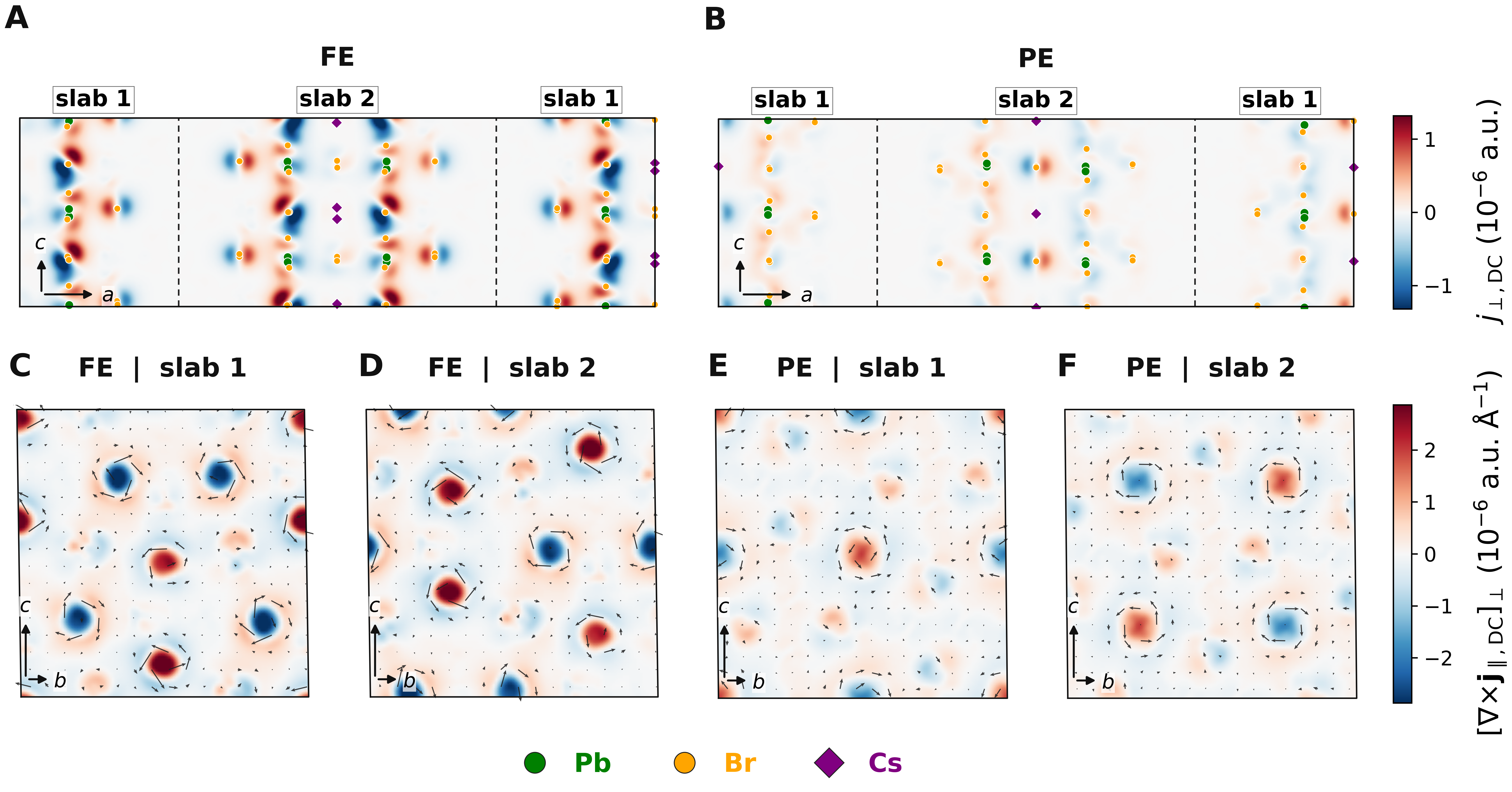}
\caption{
DC real-space current distributions in the FE and PE structures.
All quantities shown are derived from the time-averaged even-order real-space current $\mathbf j_{\mathrm{DC}}(\mathbf r)$.
(A) FE and (B) PE normal current components on the $\vec{a}$--$\vec{c}$ projection after averaging along $\vec{b}$.
In panels A and B, the red--blue background shows the current component normal to the $\vec{a}$--$\vec{c}$ plane, with red indicating current flowing into the plane and blue indicating current flowing out of the plane.
The dashed lines mark the boundaries between slabs 1 and 2 along the $\vec{a}$ direction.
See Fig.~\ref{fig:bacspbbr-structures} for the atomic structure.
Slab 1 corresponds to $[0,0.25)\cup[0.75,1)$, and slab 2 corresponds to $[0.25,0.75)$.
(C,D) FE currents averaged over slabs 1 and 2, respectively, in the $\vec{b}$--$\vec{c}$ plane.
(E,F) PE currents averaged over slabs 1 and 2, respectively, in the $\vec{b}$--$\vec{c}$ plane.
In panels C--F, black arrows indicate the in-plane current, while the red--blue background shows the signed two-dimensional curl normal to the $\vec{b}$--$\vec{c}$ plane, with opposite colors corresponding to opposite circulation directions.
Panels A and B share the same color scale, while panels C--F share a common color scale and arrow scale.
Pb, Br, and Cs atoms are shown in panels A and B, whereas the organic components are omitted.
}
\label{fig:real-space-current}
\end{figure}

Furthermore, we calculate the real-space current distributions of the relaxed ferroelectric (FE) and paraelectric (PE) structures to reveal the microscopic current in real space.
The same external field as in the finite-temperature calculations is used.
Following the procedure described above, we construct the even-order real-space current and average it over an integer number of optical periods.
The resulting time-averaged current is denoted as the DC current $\mathbf j_{\mathrm{DC}}(\mathbf r)$.
Figure~\ref{fig:real-space-current} shows the DC real-space current distributions in the FE and PE structures.
Panels A and B show pronounced spatial variations of the current normal to the $\vec{a}$--$vec{c}$ plane, with large amplitudes concentrated near Br atoms.
The slab-resolved currents in panels C--F further reveal pronounced Br-centered local current circulation.
For both FE and PE structures, these vortex currents tend to cancel each other in the total current $J_{z,\mathrm{DC}}$.
It is interesting to investigate the microscopic origin of these Br-centered vortex currents.

We decompose the total real-space current into the density-matrix--velocity ($Dv$), interband adiabatic correction, and explicit $A\Delta\rho$ components, as defined in Eq.~\eqref{eq:real-space-current-trace}.
We find that the $Dv$ component dominates the Br-centered vorticity.
By further decomposing $\mathbf j_{Dv}(\mathbf r)$ into $\mathbf j_{mn}(\mathbf r)$
$\left(\text{i.e., }\, \mathbf j_{Dv}(\mathbf r)=1/T\int\operatorname{Re}\left[\sum_{m,n}[D_{\mathrm{even}}(t)]_{nm}\,\hat{\mathbf j}_{mn}(\mathbf r,t)\right]\,\mathrm dt
\equiv\sum_{m,n}\mathbf j_{mn}(\mathbf r)\right)$,
we find that these vortex currents arise mainly from specific $(m,n)$ pairs.
By further taking into account the real-space expression in Eq.~\eqref{eq:real-space-current-trace},
\begin{equation}
\mathbf j_{mn}(\mathbf r)
=
-\frac{1}{2T}
\int
\operatorname{Re}
\Bigg\{
[D_{\mathrm{even}}(t)]_{nm}
\sum_{l\in\mathcal R}
\Big[
\phi_m^*(\mathbf r)
\phi_l(\mathbf r)
\mathbf v_{ln}(t)+
\phi_n(\mathbf r)
\phi_l^*(\mathbf r)
\mathbf v_{ml}(t)
\Big]
\Bigg\}\,dt.
\label{eq:Dv-real-space-intermediate}
\end{equation}
Here, $T$ is the duration of the time-averaging window.
Our calculations show that only a few intermediate states $l$ contribute significantly to the $\mathbf j_{mn}(\mathbf r)$ vortex currents.
We find that when $\phi_m$ and $\phi_n$ are mainly characterized by Br $p$ orbitals and $\phi_l$ is mostly characterized by a Br $s$ orbital, the overall $\mathbf j_{mn}(\mathbf r)$ vortex current is maximized.
To illustrate how this $s$--$p$ orbital coupling can generate circulation, we construct a minimal hydrogenic model with on-site $s$ and $p$ orbitals.
The analytical derivation in Sec.~\ref{si-sec:si-hydrogenic-model} of the Supplemental Information shows that such coupling can generate an atom-centered circulating current pattern.

Before we finish, we want to talk about limitations of the present CPA-NAMD-extF framework.
Since the velocity gauge and the finite number of states are used for electronic dynamics, it suffers from the common problem that some contributions from states outside the truncated adiabatic states cannot be fully described.
The finite-band correction partially compensates for missing high-energy contributions to the current but does not provide a full correction.
Moreover, it becomes worse with the ground-state adiabatic states as the basis for excitations by strong fields. In this case, the Houston basis or a dense k-point sampling will be appropriate but their computational cost will be huge for large systems.
The current method is therefore most reliable in the relatively weak field regime.

In summary, we have developed CPA-NAMD-extF for CPA-NAMD simulation with external fields. 
The method combines reduced-space density-matrix propagation with self-consistent electronic-density updates. 
Benchmarks against SALMON and PWmat validate the implementation for finite-field currents and linear optical responses. 
Calculations of 576-atom SiO$_2$ further demonstrate its applicability to large and structurally complex systems. 
Such system will be very challenging for regular rt-TDDFT calculations.
For $(\mathrm{BA})_2\mathrm{CsPb}_2\mathrm{Br}_7$, the finite-temperature second-order response shows strong fluctuations, highlighting the importance of explicit thermal sampling. 
Real-space current reveals pronounced Br-centered vortex currents in both the ferroelectric and paraelectric structures. 
Band- and orbital-resolved analysis further identifies a dominant local channel involving Br $p$ characterized states coupled through Br $s$ characterized intermediate states. 
A minimal hydrogenic model illustrates how such $s$--$p$ orbital coupling can generate atom-centered circulation. 
Overall, CPA-NAMD-extF provides an efficient framework for charge density evolution under external fields within the framework of CPA-NAMD.
\bibliography{shared/references}
\end{document}



\vspace*{-5.0em}

\begin{center}
{Supplementary Materials \par}

\vspace{1.0em}
\begin{minipage}{0.88\textwidth}
\centering
{\Large\sffamily\bfseries
Nonadiabatic molecular dynamics simulation with external field\par}
\end{minipage}

\vspace{1.5em}
{\sffamily Chengji Li\textsuperscript{\(\dagger\)} and
Fan Zheng\textsuperscript{\(*,\ddagger\)}\par}

\vspace{1.2em}
{\itshape\textsuperscript{\(\dagger\)}School of Physical Science and Technology,
ShanghaiTech University, Shanghai 201210, China\par}

\vspace{0.2em}
{\itshape\textsuperscript{\(\ddagger\)}School of Physical Science and Technology,
ShanghaiTech University, Shanghai 201210, China\par}

\vspace{0.6em}
{\sffamily E-mail: zhengfan@shanghaitech.edu.cn\par}
\end{center}

\vspace{1.0em}

\section{The CPA-NAMD-extF Method}\label{sec:si-method}

CPA-NAMD-extF propagates the electronic density matrix within an active adiabatic subspace along precomputed field-free Born--Oppenheimer molecular dynamics (BOMD) trajectories generated using Quantum ESPRESSO (QE).~\cite{giannozzi_quantum_2009,giannozzi_advanced_2017}
The electronic propagation is performed under a time-dependent external field in the velocity gauge.~\cite{yabana_time-dependent_2012,pemmaraju_velocity-gauge_2018}
With the classical-path approximation (CPA), the ions remain on the precomputed BOMD trajectory, and the field-induced electronic response is not fed back into the nuclear forces or ionic motion.~\cite{augustin_classical_1978,tully_molecular_1990,akimov_pyxaid_2013}

For each BOMD configuration $t_m$ in a trajectory, the instantaneous adiabatic Kohn-Sham (KS) orbitals are obtained by solving the ground-state Kohn--Sham equation,
\begin{equation}
H_0(t_m)
\left|\phi_n(t_m)\right\rangle
=
\varepsilon_n(t_m)
\left|\phi_n(t_m)\right\rangle .
\label{eq:ks-basis}
\end{equation}
The ground-state KS Hamiltonian matrix $H_0(t_m)$, the ground-state Hartree--exchange-correlation potential matrix $V_{0,\mathrm{Hxc}}(t_m)$, and the velocity matrix $\mathbf v(t_m)$ are then represented in the instantaneous adiabatic KS basis at $t_m$.

\pagebreak

To reduce the computational cost, the instantaneous adiabatic KS states are divided into an active subspace $\mathcal A$ and an inactive deep-valence subspace $\mathcal I$.
The active--active density-matrix block $(D_{\mathcal A \mathcal A})$ is propagated self-consistently, whereas the inactive--inactive block $(D_{\mathcal I \mathcal I})$ keeps its ground-state occupations fixed.
The time-dependent electronic density is constructed from both blocks.
The corresponding inactive charge density is constructed as
\begin{equation}
\rho_{\mathcal I}(\mathbf r;t_m)
=
\sum_{I\in\mathcal I}
f_I(t_m)
\left|
\phi_I(\mathbf r;t_m)
\right|^2.
\label{eq:rho-inactive}
\end{equation}
Here, $f_I(t_m)$ denotes the ground-state occupation of the inactive KS state $I$ at $t_m$.
The inactive subspace charge density is evaluated and stored at every BOMD time point.
Field-induced inactive--active coherences are reconstructed instantaneously for current evaluation but are not included in the self-consistent density.

To represent the active-space matrices at $t_{m+1}$ in the active adiabatic KS basis at $t_m$, we use the overlap matrix between the KS states at two adjacent BOMD configurations, defined as
\begin{equation}
U_{ab}(t_{m+1},t_m)
=
\left\langle
\phi_a(t_m)
\middle|
\phi_b(t_{m+1})
\right\rangle,
\qquad
a,b\in\mathcal A.
\label{eq:overlap}
\end{equation}
The overlap matrix $U(t_{m+1},t_m)$ is used to represent the active-space matrices at $t_{m+1}$ in the adiabatic KS basis at $t_m$. For $M\in\{H_0,V_{0,\mathrm{Hxc}},\mathbf v\}$, the transformation is given by
\begin{equation}
M^{(t_m)}(t_{m+1})
=
U(t_{m+1},t_m)
M^{(t_{m+1})}(t_{m+1})
U(t_{m+1},t_m)^{\dagger}.
\label{eq:matrix-alignment}
\end{equation}
Here, the superscript $(t_m)$ indicates the adiabatic KS basis in which the matrix is represented, and the matrices are restricted to the active subspace $\mathcal A$.
After this transformation, the active-space matrices at $t_m$ and $t_{m+1}$ are represented in the same adiabatic KS basis at $t_m$. These matrices are then linearly interpolated from the BOMD time grid onto the finer electronic propagation time grid. For $t_m\leq t\leq t_{m+1}$, the interpolation is given by
\begin{equation}
M^{(t_m)}(t)
=
M^{(t_m)}(t_m)
+
\frac{t-t_m}{t_{m+1}-t_m}
\left[
M^{(t_m)}(t_{m+1})
-
M^{(t_m)}(t_m)
\right].
\label{eq:matrix-interpolation}
\end{equation}
This procedure yields the active-space matrices $H_0(t)$, $V_{0,\mathrm{Hxc}}(t)$, and $\mathbf v(t)$ at any time $t$ within each BOMD interval, represented in the adiabatic KS basis at $t_m$.
During electronic propagation from $t_m$ to $t_{m+1}$, the active adiabatic KS basis at $t_m$ is kept fixed.
For notational simplicity, the basis superscript $(t_m)$ is omitted below when discussing propagation within a single BOMD interval.
Atomic units are used throughout unless otherwise stated.

The vector potential is related to the external electric field $\mathbf E(t)$ by
\begin{equation}
\mathbf A(t)
=
\mathbf A(t_0)
-
c\int_{t_0}^{t}
\mathbf E(t')\,dt'.
\label{eq:vector-potential}
\end{equation}
Here, $c$ is the speed of light, and $\mathbf A(t_0)=0$ is used in the present calculations.
With the sign convention adopted here, the external field potential is
\begin{equation}
V_{\mathrm{field}}(t)
=
\frac{1}{c}
\mathbf A(t)\cdot\mathbf v(t).
\label{eq:field-hamiltonian}
\end{equation}
Using the interpolated matrices $H_0(t)$, $V_{0,\mathrm{Hxc}}(t)$, and $\mathbf v(t)$, the propagation Hamiltonian is constructed as
\begin{equation}
H_{\mathrm{prop}}(t)
=
H_0(t)
-
V_{0,\mathrm{Hxc}}(t)
+
V_{\mathrm{Hxc}}[\rho(t)]
+
V_{\mathrm{field}}(t).
\label{eq:propagation-hamiltonian}
\end{equation}
Here, $H_0(t)-V_{0,\mathrm{Hxc}}(t)$ represents the part of the Hamiltonian independent of the propagated electronic density, obtained by interpolation along the BOMD trajectory.

The active-space density matrix $D_{\mathcal A\mathcal A}(t)$ is propagated self-consistently at each electronic step, following a procedure similar to that reported in Ref.\kern-0.26em\citenum{wang_efficient_2015}.
The propagation is initialized at $dt_0$ using the ground-state occupations:
\begin{equation}
D_{ab}(t_0)
=
f_a\delta_{ab},
\qquad
a,b\in\mathcal A.
\label{eq:initial-density-matrix}
\end{equation}
Here, $f_a$ denotes the ground-state occupation of the active KS state $a$. 
One BOMD step (e.g., $[t_m,t_{m+1}]$) is split into several smaller electronic steps (e.g., ($dt_n,dt_{n+1}]$).
At each electronic step from $dt_0$ to $dt_1$, an initial trial charge density $\rho_{\mathrm{tot}}^{\mathrm{trial}}(\mathbf r,dt_1)$ at $dt_1$ is extrapolated from the converged charge densities of previous electronic steps.
This trial density is used to construct the propagation Hamiltonian at $dt_1$ according to Eq.~\eqref{eq:propagation-hamiltonian}.
The propagation Hamiltonian at time $t\in[dt_0,dt_1]$ is then linearly interpolated within the electronic step.
The active-space density matrix is then propagated as
\begin{equation}
i\frac{dD_{\mathcal A\mathcal A}(t)}{dt}
=
\left[
H_{\mathrm{prop}}(t),
D_{\mathcal A\mathcal A}(t)
\right].
\label{eq:density-matrix}
\end{equation}
This equation is integrated using a fourth-order Runge--Kutta method.
The propagated density matrix is then used to reconstruct the active-space electronic density:
\begin{equation}
\rho_{\mathcal A}(\mathbf r,t_1)
=
\sum_{a,b\in\mathcal A}
\phi_a(\mathbf r;t_m)
D_{ab}(t_1)
\phi_b^*(\mathbf r;t_m).
\label{eq:rho-active}
\end{equation}
The total electronic density is constructed by adding the inactive deep-valence contribution:
\begin{equation}
\rho_{\mathrm{tot}}(\mathbf r,dt_1)
=
\rho_{\mathcal A}(\mathbf r,dt_1)
+
\rho_{\mathcal I}(\mathbf r;t_m).
\label{eq:rho-calc}
\end{equation}
Here, $\rho_{\mathcal I}(\mathbf r;t_m)$ is defined in Eq.~\eqref{eq:rho-inactive} and remains fixed within each BOMD interval.
The reconstructed density $\rho_{\mathrm{tot}}(\mathbf r,dt_1)$ is compared with the trial density $\rho_{\mathrm{tot}}^{\mathrm{trial}}(\mathbf r,dt_1)$.
If the prescribed convergence criterion is satisfied (i.e., $\left\lVert \rho_{\mathrm{tot}}(\mathbf r,dt_1)-\rho_{\mathrm{tot}}^{\mathrm{trial}}(\mathbf r,dt_1)\right\rVert \leq \epsilon$), the reconstructed density and the propagated density matrix are accepted as the converged values.
Otherwise, the two densities are linearly mixed:
\begin{equation}
\rho_{\mathrm{tot}}^{\mathrm{new}}(\mathbf r,dt_1)
=
(1-\alpha)
\rho_{\mathrm{tot}}^{\mathrm{trial}}(\mathbf r,dt_1)
+
\alpha
\rho_{\mathrm{tot}}(\mathbf r,dt_1).
\label{eq:rho-mixing}
\end{equation}
Here, $\alpha$ is the density-mixing parameter.
The mixed density is used as the trial density for the next self-consistent iteration.
The density-matrix propagation and density mixing are repeated until the prescribed convergence criterion is satisfied.
This self-consistent procedure yields the time-dependent density matrix $D_{\mathcal A\mathcal A}(t)$ and electronic density $\rho(\mathbf r,t)$ at each electronic step.

At the boundary between two adjacent BOMD intervals, the converged active-space density matrix is transformed from the adiabatic KS basis at $t_m$ to that at $t_{m+1}$.
Such transformation is given by
\begin{equation}
D_{\mathcal A\mathcal A}^{(t_{m+1})}(t_{m+1})
=
U^\dagger(t_{m+1},t_m)
D_{\mathcal A\mathcal A}^{(t_m)}(t_{m+1})
U(t_{m+1},t_m).
\label{eq:density-matrix-boundary}
\end{equation}
Here, $U(t_{m+1},t_m)$ is the overlap matrix defined in Eq.~\eqref{eq:overlap}.
The transformed density matrix serves as the initial condition for electronic propagation in the next BOMD interval.

After convergence of the active-space self-consistent propagation, the current is evaluated using an effective density matrix in the combined space $\mathcal R=\mathcal I\oplus\mathcal A$.
The density matrix is assembled as
\begin{equation}
D(t)
=
\begin{pmatrix}
D_{\mathcal I\mathcal I}(t_m) & D_{\mathcal I\mathcal A}(t)\\
D_{\mathcal A\mathcal I}(t) & D_{\mathcal A\mathcal A}(t)
\end{pmatrix}.
\label{eq:retained-blocks}
\end{equation}
The active-space density matrix $D_{\mathcal A\mathcal A}(t)$ is obtained from the self-consistent propagation, while $D_{\mathcal I\mathcal I}(t_m)$ contains the frozen ground-state occupations $f_I(t_m)$ which are fixed.
The inactive--active coherence $D_{\mathcal I\mathcal A}(t)$ is reconstructed instantaneously for current evaluation, with $D_{\mathcal A\mathcal I}(t)=D_{\mathcal I\mathcal A}^{\dagger}(t)$.
These off-diagonal blocks are not included in the self-consistent electronic density.

To calculate $D_{\mathcal I\mathcal A}(t)$, we introduce the Hamiltonian in the subspace $\mathcal R$, written in block form as
\begin{equation}
H^{\mathcal R}(t)
=
\begin{pmatrix}
H_{\mathcal I\mathcal I}(t) & H_{\mathcal I\mathcal A}(t)\\
H_{\mathcal A\mathcal I}(t) & H_{\mathcal A\mathcal A}(t)
\end{pmatrix}.
\label{eq:retained-hamiltonian}
\end{equation}
Here, $H_{\mathcal A\mathcal A}(t)=H_{\mathrm{prop}}(t)$ is the self-consistent active-space propagation Hamiltonian.
The inactive-space Hamiltonian is approximated by its diagonal ground-state form, $H_{IJ}(t)\simeq\varepsilon_I(t_m)\delta_{IJ}$.
The inactive--active coupling is approximated by the external-field contribution, $H_{\mathcal I\mathcal A}(t)\simeq V_{\mathrm{field},\mathcal I\mathcal A}(t)=\mathbf A(t)\cdot\mathbf v_{\mathcal I\mathcal A}(t)/c$, neglecting the time-dependent Hartree--exchange-correlation contribution to this block.
The corresponding active--inactive block satisfies $H_{\mathcal A\mathcal I}(t)=H_{\mathcal I\mathcal A}^{\dagger}(t)$.

Using Eqs.~\eqref{eq:retained-hamiltonian} and \eqref{eq:retained-blocks}, together with the approximations described above, the $\mathcal I\mathcal A$ block of the Liouville equation becomes
\begin{equation}
i\frac{dD_{\mathcal I\mathcal A}(t)}{dt}
=
H_{\mathcal I\mathcal I}(t)D_{\mathcal I\mathcal A}(t)
-D_{\mathcal I\mathcal A}(t)H_{\mathrm{prop}}(t)
+S_{\mathcal I\mathcal A}(t).
\label{eq:dia-liouville}
\end{equation}
Here, $S_{\mathcal I\mathcal A}(t)$ is defined as
\begin{equation}
S_{\mathcal I\mathcal A}(t)
=V_{\mathrm{field},\mathcal I\mathcal A}(t)D_{\mathcal A\mathcal A}(t)-D_{\mathcal I\mathcal I}(t_m)V_{\mathrm{field},\mathcal I\mathcal A}(t).
\label{eq:dia-source}
\end{equation}
In the following, the time derivative of $D_{\mathcal I\mathcal A}(t)$ is neglected.
Equation~\eqref{eq:dia-liouville} then reduces to
\begin{equation}
0\simeq
S_{\mathcal I\mathcal A}(t)
+H_{\mathcal I\mathcal I}(t)D_{\mathcal I\mathcal A}(t)
-D_{\mathcal I\mathcal A}(t)H_{\mathrm{prop}}(t).
\label{eq:dia-adiabatic}
\end{equation}
To solve this algebraic equation, we diagonalize the converged active-space propagation Hamiltonian using its unitary eigenvector matrix $W(t)$:
\begin{equation}
W^\dagger(t)H_{\mathrm{prop}}(t)W(t)
=
\operatorname{diag}\left[\widetilde{\varepsilon}_A(t)\right].
\label{eq:dia-diagonalization}
\end{equation}
Here, $\widetilde{\varepsilon}_A(t)$ denotes the instantaneous state-$A$ eigenvalue of the active-space propagation Hamiltonian.
The inactive--active density-matrix block and its source term are transformed into this eigenbasis as $\widetilde{D}_{\mathcal I\mathcal A}(t)=D_{\mathcal I\mathcal A}(t)W(t)$ and $\widetilde{S}_{\mathcal I\mathcal A}(t)=S_{\mathcal I\mathcal A}(t)W(t)$.
Since $H_{\mathcal I\mathcal I}(t)$ is approximated by its diagonal ground-state form, Eq.~\eqref{eq:dia-adiabatic} can be solved elementwise as
\begin{equation}
\widetilde{D}_{IA}(t)
=
-\frac{
\widetilde{S}_{IA}(t)
}{
\varepsilon_I(t_m)-\widetilde{\varepsilon}_A(t)
}.
\label{eq:dia-instantaneous}
\end{equation}
The reconstructed coherence is transformed back to the adiabatic KS basis at $t_m$ using $D_{\mathcal I\mathcal A}(t)=\widetilde{D}_{\mathcal I\mathcal A}(t)W^\dagger(t)$ for current evaluation.
Although not explicitly propagated, $D_{\mathcal I\mathcal A}(t)$ remains time dependent through the external potential and active-space eigenstates.

With the reconstructed density matrix $D(t)$, the macroscopic cell-averaged current density is calculated as $\mathbf J(t)=1/\Omega \, \mathbf j(t)$, and
\begin{equation}
\mathbf j(t)
=
\mathbf j^{\mathrm{para}}(t)
+
\mathbf j^{\mathrm{dia}}(t)
+
\mathbf j^{\mathrm{corr}}(t),
\label{eq:current}
\end{equation}
where $\Omega$ is the unit-cell volume.
The paramagnetic contribution is evaluated as
\begin{equation}
\mathbf j^{\mathrm{para}}(t)
=
-\operatorname{Re}
\left\{
\operatorname{Tr}
\left[
D(t)\mathbf v(t)
\right]
\right\}.
\label{eq:current-paramagnetic}
\end{equation}
Here, the trace is taken over the subspace $\mathcal R$, including the active--active, inactive--inactive, and reconstructed inactive--active contributions.
The diamagnetic contribution is $\mathbf j^{\mathrm{dia}}(t)=-N_e\mathbf A(t)/c$, where $N_e$ is the number of electrons in the unit cell.

In the velocity gauge, current convergence generally requires a large number of high-energy unoccupied states.
To partially compensate for the missing contributions caused by finite-band truncation, we employ the first-order adiabatic correction derived in Ref.~\citenum{yakovlev_adiabatic_2017}.
The correction is given by
\begin{equation}
\mathbf j^{\mathrm{corr}}(t)
=
\frac{1}{c}
\sum_{n\in\mathrm{occ}}
\left[
\mathbf A(t)
-
2
\sum_{\substack{i=1\\i\neq n}}^{N_{\mathrm b}}
\frac{
\operatorname{Re}
\left[
\left(
\mathbf v_{ni}(t)\cdot\mathbf A(t)
\right)
\mathbf v_{in}(t)
\right]
}{
\varepsilon_i(t_m)-\varepsilon_n(t_m)
}
\right].
\label{eq:current-correction}
\end{equation}
Here, $N_{\mathrm b}$ is the number of bands retained in the current evaluation, and the sum over $n$ runs over the occupied ground-state KS states.
The adiabatic energies $\varepsilon_n(t_m)$ are taken from the ground-state KS calculations at each BOMD configuration, whereas the velocity matrix elements $\mathbf v_{ni}(t)$ are interpolated onto the electronic propagation time grid.

\section{Current Post-Processing}\label{sec:si-response}

The macroscopic current density $\mathbf J(t)$ is evaluated according to Eq.~\eqref{eq:current}.
For linear optical responses, the paramagnetic and finite-band correction currents are used for spectral analysis, while nonlinear responses are extracted from the total current.

\subsection{Linear-Response Analysis}\label{sec:si-linear-response}

For linear-response calculations, a weak $\delta$-kick perturbation is applied in the velocity gauge.
Using the relation between the electric field and vector potential defined in Eq.~\eqref{eq:vector-potential}, the $\delta$-kick is introduced through a step-like vector potential:
\begin{equation}
\frac{\mathbf A(t)}{c}
=
-E_{\mathrm{kick}}
\Theta(t-t_0)
\hat{\mathbf e},
\label{eq:kick-vector-potential}
\end{equation}
which corresponds to an electric field
\begin{equation}
\mathbf E(t)
=
E_{\mathrm{kick}}
\delta(t-t_0)
\hat{\mathbf e}.
\label{eq:kick-electric-field}
\end{equation}
Here, $E_{\mathrm{kick}}$ is the time-integrated kick strength, $t_0$ is the kick time, $\hat{\mathbf e}$ denotes the field-polarization direction, and $\Theta$ is the step function.
The kick strength is chosen sufficiently small to ensure a linear response.

The paramagnetic and finite-band correction currents are used for spectral analysis.
The current projected along the field-polarization direction is defined as
\begin{equation}
J_{\parallel}(t)
=
\frac{1}{\Omega}
\hat{\mathbf e}\cdot
\left[
\mathbf j^{\mathrm{para}}(t)
+
\mathbf j^{\mathrm{corr}}(t)
\right].
\label{eq:linear-projected-current}
\end{equation}
The zero-field reference is subtracted to obtain the induced current $\Delta J_{\parallel}(t)=J_{\parallel}(t)-J_{\parallel}(t;0)$, where $J_{\parallel}(t;0)$ is taken as the current just before the application of such kick field.
To suppress spectral oscillations caused by a finite-time propagation, a Gaussian damping window is applied:
\begin{equation}
w(t)
=
\exp\left[
-B
\left(
\frac{t-t_{\min}}{T}
\right)^2
\right],
\qquad
T=t_{\max}-t_{\min},
\label{eq:time-window}
\end{equation}
where $B$ is the broadening parameter, and $t_{\min}$ and $t_{\max}$ define the time interval used for spectral analysis.
A zero-frequency offset associated with the step-like vector potential is also removed.
The processed current is constructed as
\begin{equation}
\Delta J_{\parallel}^{\mathrm{proc}}(t)
=
\left[
\Delta J_{\parallel}(t)
-
\gamma\frac{A_{\parallel}(t)}{c}
\right]
w(t),
\label{eq:processed-current}
\end{equation}
where
\begin{equation}
A_{\parallel}(t)
=
\hat{\mathbf e}\cdot\mathbf A(t),
\qquad
\gamma
=
\frac{
\displaystyle
\int_{t_{\min}}^{t_{\max}}
\Delta J_{\parallel}(t)w(t)\,dt
}{
\displaystyle
\int_{t_{\min}}^{t_{\max}}
[A_{\parallel}(t)/c]w(t)\,dt
}.
\label{eq:dc-correction}
\end{equation}
The coefficient $\gamma$ is chosen so that the processed current has a vanishing zero-frequency component within the analysis window.

The Fourier transform is defined as
\begin{equation}
F(\omega)
=
\int dt\,F(t)e^{-i\omega t}.
\label{eq:fourier-transform}
\end{equation}
Under this convention, the Fourier transform of the $\delta$-kick electric field is
\begin{equation}
E_{\parallel}(\omega)
=
E_{\mathrm{kick}}e^{-i\omega t_0}.
\label{eq:kick-field-frequency}
\end{equation}
The raw imaginary part of the dielectric function is obtained from the Fourier-transformed current:
\begin{equation}
\begin{aligned}
\varepsilon_2^{\mathrm{raw}}(\omega)
&=
-\operatorname{Im}
\left[
\frac{
\Delta J_{\parallel}^{\mathrm{proc}}(\omega)
}{
i\omega E_{\parallel}(\omega)
}
\right]\\
&=
\frac{1}{\omega}
\operatorname{Re}
\left[
\frac{
\Delta J_{\parallel}^{\mathrm{proc}}(\omega)
}{
E_{\parallel}(\omega)
}
\right].
\end{aligned}
\label{eq:epsilon-imag-raw}
\end{equation}
An effective $f$-sum normalization is applied to determine the absolute spectral weight:
\begin{equation}
\varepsilon_2(\omega)
=
\frac{
2\pi^2N_e/\Omega
}{
\displaystyle
\int_{\omega_{\min}}^{\omega_{\max}}
\omega\varepsilon_2^{\mathrm{raw}}(\omega)\,d\omega}
\varepsilon_2^{\mathrm{raw}}(\omega),
\label{eq:f-sum-normalization}
\end{equation}
Here, $[\omega_{\min},\omega_{\max}]$ denotes the frequency range used for normalization.
This procedure rescales the raw spectrum so that its integrated spectral weight matches the effective $f$-sum spectral weight corresponding to $N_e$ electrons.

The real part of the dielectric function is obtained through the Kramers--Kronig relation:~\cite{kuzmenko_kramerskronig_2005}
\begin{equation}
\varepsilon_1(\omega)
=
1+
\frac{2}{\pi}
\mathcal P
\int_0^\infty
\frac{
\omega'\varepsilon_2(\omega')
}{
\omega'^2-\omega^2
}
\,d\omega'.
\label{eq:epsilon-real}
\end{equation}
Here, $\mathcal P$ denotes the Cauchy principal value.
In practice, the integral is evaluated over the available finite frequency window.

The absorption coefficient is then calculated as
\begin{equation}
\alpha(E_{\mathrm{ph}})
=
\frac{4\pi E_{\mathrm{ph}}}{hc}
\sqrt{
\frac{
\sqrt{
\varepsilon_1^2(E_{\mathrm{ph}})
+
\varepsilon_2^2(E_{\mathrm{ph}})
}
-
\varepsilon_1(E_{\mathrm{ph}})
}{2}
},
\label{eq:absorption}
\end{equation}
where $E_{\mathrm{ph}}=\hbar\omega$ is the photon energy.

\subsection{Nonlinear-Response Analysis}\label{sec:si-nonlinear-response}

For nonlinear-response calculations, finite-amplitude oscillatory fields are applied.
The monochromatic perturbation is introduced through the vector potential
\begin{equation}
\mathbf A(t)
=
A_0f(t)
\cos\left[
\omega_0(t-t_{\mathrm{on}})+\phi
\right]
\hat{\mathbf e},
\label{eq:monochromatic-vector-potential}
\end{equation}
where $A_0$ is the vector-potential amplitude, $f(t)$ is a $\sin^2$-shaped turn-on envelope that smoothly increases from zero to unity, $t_{\mathrm{on}}$ is the field-onset time, $\omega_0$ is the optical frequency, and $\phi$ is the phase.
In the constant-envelope region where $f(t)=1$, the electric-field amplitude is
\begin{equation}
E_0
=
\frac{\omega_0A_0}{c}.
\label{eq:field-amplitude}
\end{equation}

The field-induced even-order current is isolated from the positive-field, negative-field, and zero-field calculations as
\begin{equation}
\mathbf J_{\mathrm{even}}(t)
=
\frac{
\mathbf J(t;+E_0)
+
\mathbf J(t;-E_0)
}{2}
-
\mathbf J(t;0).
\label{eq:si-even-order-current}
\end{equation}
Here, $\mathbf J(t;0)$ removes the field-independent background, while the positive- and negative-field combination cancels all odd-order contributions.
In the weak-field regime,
\begin{equation}
\mathbf J_{\mathrm{even}}(t)
=
E_0^2\mathbf J^{[2]}(t)
+
\mathcal O(E_0^4),
\label{eq:si-even-order-expansion}
\end{equation}
so the extracted current is dominated by the second-order response.

For a monochromatic field at frequency $\omega_0$, the DC second-order response corresponds to the zero-frequency component of the even-order current.
This component is obtained by averaging over a selected time window:
\begin{equation}
\mathbf J_{\mathrm{DC}}
=
\frac{1}{t_b-t_a}
\int_{t_a}^{t_b}
\mathbf J_{\mathrm{even}}(t)\,dt.
\label{eq:si-dc-current}
\end{equation}
The corresponding second-order conductivity is calculated as
\begin{equation}
\sigma^{(2)}(0;\omega_0,-\omega_0)
=
\frac{\mathbf J_{\mathrm{DC}}}{E_0^2}.
\label{eq:si-second-order-conductivity}
\end{equation}
Here, $t_a$ and $t_b$ denote the beginning and end of the averaging window.
The window is chosen within the constant-envelope region to span an integer number of optical periods, $t_b-t_a=N(2\pi/\omega_0)$, where $N$ is an integer.

\section{Real-Space Current}\label{sec:si-real-space-current}

To construct the real-space current density, we first express the finite-band adiabatic correction in a density-matrix form.
Starting from Eq.~\eqref{eq:current-correction}, the off-diagonal contribution to the macroscopic current density is
\begin{equation}
\mathbf J^{\mathrm{corr},\mathrm{off}}(t)
=
-\frac{2}{c\Omega}
\sum_{\substack{m,n\in\mathcal R\\m\ne n}}
\frac{f_n}{\varepsilon_m(t_m)-\varepsilon_n(t_m)}
\operatorname{Re}
\left[
\left(
\mathbf A(t)\cdot\mathbf v_{nm}(t)
\right)
\mathbf v_{mn}(t)
\right].
\label{eq:realspace-current-correction-offdiagonal}
\end{equation}
Here, $f_n$ denotes the ground-state occupation of state $n$, following the occupation convention used in Eq.~\eqref{eq:current-correction}.
Note that since the conduction band has $f_n=0$, the summation is expanded to subspace $\mathcal{R}$.
The occupation-sum part $1/c \, \sum_{n \in \mathrm{occ}} \mathbf{A}(t)$ is left for further evaluation.
This equation can be further simplified owing to the real-valued $\mathbf{A}$ and hermiticity of $\mathbf{v}$.
\begin{equation}
\begin{gathered}
2\sum_{m\ne n}
\frac{f_n}{\varepsilon_m(t_m)-\varepsilon_n(t_m)}
\operatorname{Re}\left[
(\mathbf{A}(t)\cdot\mathbf{v}_{mn}(t))
\mathbf{v}_{nm}(t)
\right]
\\[3pt]
=
\sum_{m\ne n}
\frac{f_n-f_m}{\varepsilon_m(t_m)-\varepsilon_n(t_m)}
\operatorname{Re}\left[
(\mathbf{A}(t)\cdot\mathbf{v}_{mn}(t))
\mathbf{v}_{nm}(t)
\right].
\end{gathered}
\label{eq:realspace-current-pair-symmetrization}
\end{equation}
This identity holds for the complete double sum rather than for individual matrix elements.

Using this identity, we introduce the correction matrix
\begin{equation}
\left[\delta D_c(t)\right]_{mn}
=
\frac{f_n-f_m}
{\varepsilon_m(t_m)-\varepsilon_n(t_m)}
\frac{
\mathbf A(t)\cdot\mathbf v_{mn}(t)
}{c},
\qquad m\ne n.
\label{eq:realspace-delta-density-correction}
\end{equation}
The diagonal elements are set to zero.
This is because the diagonal elements of $\mathbf{v}$ vanish, and the following trace only involves off-diagonal matrix elements.
For numerically degenerate states, the corresponding matrix elements are also set to zero, with the same pairs excluded from the original correction sum.
The resulting matrix $\delta D_c(t)$ is Hermitian.
Substituting Eq.~\eqref{eq:realspace-current-pair-symmetrization} into Eq.~\eqref{eq:realspace-current-correction-offdiagonal}, we obtain
\begin{equation}
\begin{aligned}
\mathbf J^{\mathrm{corr},\mathrm{off}}(t)
&=
-\frac{1}{c\Omega}
\sum_{m\ne n}
\frac{f_n-f_m}
{\varepsilon_m(t_m)-\varepsilon_n(t_m)}
\operatorname{Re}\left[
(\mathbf{A}(t)\cdot\mathbf{v}_{mn}(t))
\mathbf{v}_{nm}(t)
\right]\\
&=
-\frac{1}{\Omega}
\operatorname{Re}
\sum_{m,n\in\mathcal R}
\left[\delta D_c(t)\right]_{mn}
\mathbf v_{nm}(t)\\
&=
-\frac{1}{\Omega}
\operatorname{Re}\operatorname{Tr}
\left[
\delta D_c(t)\mathbf v(t)
\right].
\end{aligned}
\label{eq:realspace-current-correction-matrix}
\end{equation}
Thus, $\delta D_c(t)$ provides an equivalent matrix representation of the off-diagonal finite-band correction.
It is used only for current evaluation and does not enter the self-consistent electronic-density update.

We next derive the real-space current density.
Using the position-resolved current operator $\hat{\mathbf j}(\mathbf r)$ defined in the main text and inserting the adiabatic KS states, we express the contribution from $D(t)+\delta D_c(t)$ as
\begin{equation}
\begin{aligned}
\operatorname{Tr}
\left[
\left(D(t)+\delta D_c(t)\right)
\hat{\mathbf j}(\mathbf r)
\right] &=
\operatorname{Re}
\sum_{m,n\in\mathcal R}
\left[D(t)+\delta D_c(t)\right]_{nm}
\langle m|\hat{\mathbf j}(\mathbf r)|n\rangle\\
&=
-\operatorname{Re}
\sum_{n,m\in\mathcal R}
\left[D(t)+\delta D_c(t)\right]_{nm}
\sum_{l\in\mathcal R}
\phi_l(\mathbf r;t_m)
\mathbf v_{ln}(t)
\phi_m^*(\mathbf r;t_m).
\end{aligned}
\label{eq:realspace-current-orbital-expansion}
\end{equation}
Here, $\phi_n(\mathbf r;t_m)$ denotes the adiabatic KS orbital at the current BOMD configuration $t_m$.
The local diamagnetic contribution is $-\mathbf A(t)\rho(\mathbf r,t)/c$. 
Here, $\rho(\mathbf r,t)$ is the self-consistent electronic density defined in Eq.~\eqref{eq:rho-calc}, constructed from the propagated active-space density matrix and the frozen inactive-state contribution. 
The reconstructed inactive--active coherences are included in $D(t)$ for current evaluation but do not contribute to $\rho(\mathbf r,t)$. 
The occupation-sum part of Eq.~\eqref{eq:current-correction} is represented positionally by $\mathbf A(t)\rho_0(\mathbf r;t)/c$, where $\rho_0(\mathbf r;t)$ is the ground-state reference density.
Combining these contributions gives the explicit $\mathbf A(t)[\rho_0(\mathbf r;t)-\rho(\mathbf r,t)]/c$ term in the real-space current density.

To evaluate the real-space current density, we discretize the spatial coordinates onto the DFT FFT grid ${\mathbf r_g}$ with $N_g$ points.
At each grid point $\mathbf r_g$, Eq.~\eqref{eq:realspace-current-orbital-expansion} gives
\begin{equation}
\begin{aligned}
\mathbf j(\mathbf r_g,t)
={}&
-\operatorname{Re}
\sum_{l,n,m\in\mathcal R}
\phi_l(\mathbf r_g;t_m)
\mathbf v_{ln}(t)
\left[D(t)+\delta D_c(t)\right]_{nm}
\phi_m^*(\mathbf r_g;t_m)\\
&+
\frac{\mathbf A(t)}{c}
\left[
\rho_0(\mathbf r_g;t)
-
\rho(\mathbf r_g,t)
\right].
\end{aligned}
\label{eq:realspace-current-density}
\end{equation}
The first term contains the density-matrix--velocity ($Dv$) contribution and the off-diagonal finite-band correction, while the second combines the diamagnetic contribution and the occupation-sum correction.
Using the normalization of the KS orbitals and the electronic densities, the spatial average gives
\begin{equation}
\begin{aligned}
\frac{1}{\Omega} \int \mathbf j(\mathbf r,t) \,\mathrm d^3\mathbf r 
={}&
-\frac{1}{\Omega}
\operatorname{Re}\operatorname{Tr}
\left[
\left(D(t)+\delta D_c(t)\right)
\mathbf v(t)
\right]+
\frac{\mathbf A(t)}{c\Omega}
\left[
\sum_{n\in\mathcal R}f_n-N_e
\right]\\
={}&
\mathbf J(t).
\end{aligned}
\label{eq:realspace-current-volume-average}
\end{equation}
Thus, the spatially averaged real-space current reproduces the macroscopic current, including the finite-band correction.

The even-order real-space current is obtained using the same positive- and negative-field combination as in Eq.~\eqref{eq:si-even-order-current}:
\begin{equation}
\mathbf j_{\mathrm{even}}(\mathbf r,t)
=
\frac{
\mathbf j(\mathbf r,t;+E_0)
+
\mathbf j(\mathbf r,t;-E_0)
}{2}
-
\mathbf j(\mathbf r,t;0).
\label{eq:realspace-even-current}
\end{equation}
The DC real-space current is then calculated by averaging over the selected time window,
\begin{equation}
\mathbf j_{\mathrm{DC}}(\mathbf r)
=
\frac{1}{t_b-t_a}
\int_{t_a}^{t_b}
\mathbf j_{\mathrm{even}}(\mathbf r,t)\,dt.
\label{eq:realspace-dc-current}
\end{equation}
Its spatial average reproduces the macroscopic DC current defined in Eq.~\eqref{eq:si-dc-current}:
\begin{equation}
\mathbf J_{\mathrm{DC}}
=
\frac{1}{\Omega}
\int
\mathbf j_{\mathrm{DC}}(\mathbf r)\,\mathrm d^3\mathbf r.
\label{eq:realspace-dc-volume-average}
\end{equation}

\section{Real-Space Current Decomposition and Hydrogenic Model}\label{sec:si-hydrogenic-model}

To investigate the microscopic origin of the Br-centered current vortices shown in Fig.~\ref{fig:real-space-current}, we analyze the time-averaged even-order real-space current in the zero-temperature ferroelectric (FE) and paraelectric (PE) structures.
The DC current maps are obtained by averaging the even-order real-space current over an integer number of optical periods.

For the spatial visualization shown in Fig.~\ref{fig:real-space-current}A and B, the current is averaged along the $\vec{b}$ axis and projected onto the $\vec{a}$--$\vec{c}$ plane.
The structure is further divided into two slabs along the $\vec{a}$ direction, with fractional-coordinate intervals $[0,0.25)\cup[0.75,1)$ and $[0.25,0.75)$.
The corresponding slab-averaged currents are projected onto the $\vec{b}$--$\vec{c}$ plane shown in Fig.~\ref{fig:real-space-current}C--F. 
The signed two-dimensional vorticity normal to each projection plane is calculated as
\begin{equation}
W_{bc}(\mathbf r)
=
\frac{\partial j_c(\mathbf r)}{\partial u_b}
-
\frac{\partial j_b(\mathbf r)}{\partial v_c},
\label{eq:vortex-two-dimensional-curl}
\end{equation}
where $u_b$ and $v_c$ are two orthogonal directions along $\vec{b}$ and $\vec{c}$, respectively.
The spatial derivatives are evaluated on the FFT grid.
The resulting vorticity maps characterize the local circulation patterns shown in Fig.~\ref{fig:real-space-current}C--F.

According to Eq.~\eqref{eq:realspace-current-density}, the real-space current consists of the density-matrix--velocity ($Dv$), off-diagonal finite-band correction ($\delta D_c$ related terms), and explicit field--density contributions ($\mathbf{A}(t)/c \, [\rho_0(\mathbf{r},t_m)-\rho(\mathbf{r},t)]$).
The off-diagonal finite-band correction cancels in the even-order response because it is linear in the external field.
For a representative current field of the FE structure, the $Dv$ term is the dominant one.
We focus only on the $Dv$ contribution in the following microscopic analysis.
The even-order $Dv$ current is evaluated as
\begin{equation}
\begin{aligned}
\mathbf j_{\mathrm{even},Dv}(\mathbf r,t)
=
-\operatorname{Re}
\sum_{l,n,m\in\mathcal R}
\phi_l(\mathbf r;t_m)
\mathbf v_{ln}(t)
[D_{\mathrm{even}}(t)]_{nm}
\phi_m^*(\mathbf r;t_m),
\end{aligned}
\label{eq:realspace-even-current-explicit}
\end{equation}
where $D_{\mathrm{even}}(t)=[D(t;+E_0)+D(t;-E_0)]/2-D(t;0)$.

\begin{figure}[!t]
\centering
\includegraphics[width=\linewidth]{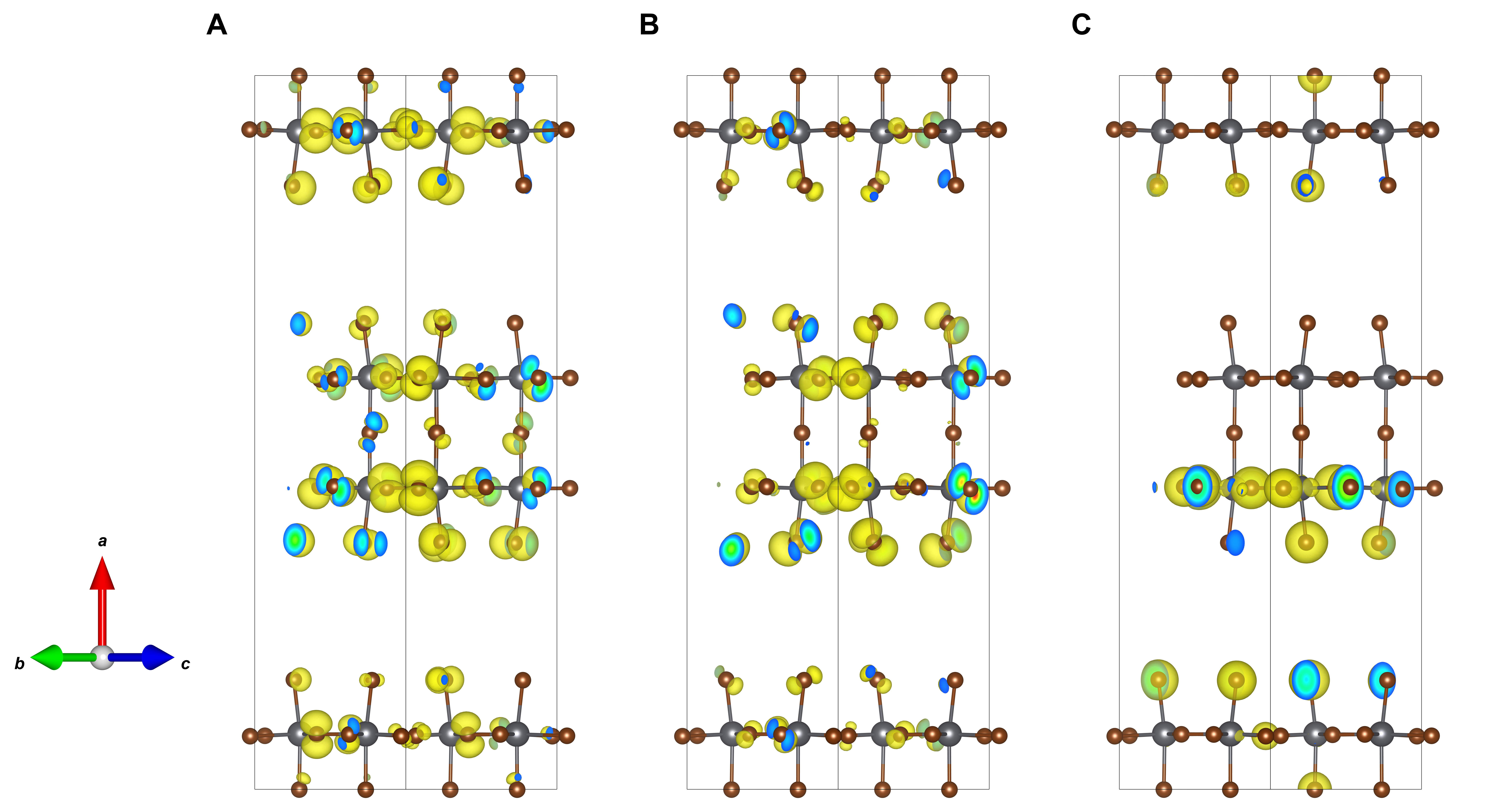}
\caption{
Real-space probability-density isosurfaces of the representative FE states.
(A,B) States $m=535$ and $n=543$ forming the selected band pair.
(C) The leading intermediate state $l=149$.
Only the Pb--Br framework is shown, and the crystallographic axes are indicated at the lower left.
}
\label{fig:representative-state-densities}
\end{figure}

To identify the electronic states associated with the Br-centered current vortices, we decompose the even-order $Dv$ current into contributions from individual pairs of adiabatic KS states.
\begin{equation}
\begin{aligned}
\overline{\mathbf j}_{\mathrm{DC},Dv}(\mathbf r)
= &
-\frac{1}{t_b-t_a}
\int_{t_a}^{t_b}
\operatorname{Re}
\left\{
\sum_{m,n \in \mathcal{R}} [D_{\mathrm{even}}(t)]_{nm}
\sum_{l \in \mathcal{R}}
\left[
\phi_l(\mathbf r;t_m)
\mathbf v_{ln}(t)
\phi_m^*(\mathbf r;t_m)
\right]
\right\}\,dt \\
&\equiv
-\frac{1}{t_b-t_a}
\int_{t_a}^{t_b}
\sum_{m,n \in \mathcal{R}}
\mathbf j_{\mathrm{DC},Dv}^{(m,n)}(\mathbf r,t)\,dt
\\
&\equiv
-\sum_{m,n\in\mathcal R}
\overline{\mathbf j}_{\mathrm{DC},Dv}^{(m,n)}(\mathbf r).
\end{aligned}
\label{eq:vortex-pair-current}
\end{equation}
where we define the band-pair contribution as
\begin{equation}
\mathbf j_{\mathrm{DC},Dv}^{(m,n)}(\mathbf r,t)
\equiv
\frac{1}{2}
\operatorname{Re}
\left\{
[D_{\mathrm{even}}(t)]_{mn}
\sum_{l\in\mathcal R}
\mathbf K_{mn}^l(\mathbf r,t)
\right\},
\label{eq:vortex-pair-contribution}
\end{equation}
with
\begin{equation}
\mathbf K_{mn}^l(\mathbf r,t)
=
\phi_n^*(\mathbf r;t_m)
\phi_l(\mathbf r;t_m)
\mathbf v_{lm}(t)
+
\phi_m(\mathbf r;t_m)
\phi_l^*(\mathbf r;t_m)
\mathbf v_{nl}(t).
\label{eq:vortex-intermediate-kernel}
\end{equation}
For the band-pair analysis, the time average $\overline{\mathbf j}_{\mathrm{DC},Dv}^{(m,n)}(\mathbf r)$ is evaluated over 16 complete optical cycles from $3.0$ to $19.5~\mathrm{fs}$ using piecewise-linear integration of the saved current samples.
By checking all $(m,n)$ pairs of $\overline{\mathbf j}_{\mathrm{DC},Dv}^{(m,n)}(\mathbf r)$ for their contributions of vorticity, the highest-ranked FE pair, $(m,n)=(535,543)$ is selected as an example.
Other contributing pairs show a similar mechanism, as discussed later.
To further identify the orbital channels associated with these band pairs, we decompose the time-averaged current into contributions from each intermediate state $l$.
Since the density-matrix element $[D_{\mathrm{even}}(t)]_{mn}$ is only a complex number, the vorticity of currents comes completely from the KS states and velocity matrix.
For the representative FE pair $(m,n)=(535,543)$, we find that state $l=149$ contributes most to the vorticity.
Atomic-orbital projections show that states 535 and 543 are predominantly of Br $4p$ character, whereas state 149 is predominantly of Br $4s$ character.
The corresponding real-space probability densities are shown in Fig.~\ref{fig:representative-state-densities}.
Other state-$m,n$ with $p$ orbital character and state-$l$ with $s$ orbital character also introduce significant vortex current.
This suggests that the local Br $4s$--$4p$ velocity coupling is an important microscopic channel associated with the Br-centered current vortices.

To further illustrate how local $s$--$p$ velocity coupling can generate an azimuthal current kernel, we construct a minimal hydrogenic model. The model describes the spatial geometry of the current kernel rather than the quantitative electronic structure or current magnitude of Br.
Working in atomic units and neglecting the nonlocal pseudopotential contribution, the velocity operator is
\begin{equation}
\hat{\mathbf v}
=
-i\nabla.
\label{eq:hydrogen-velocity-operator}
\end{equation}
We consider a $1s$ intermediate state and two orthogonal $2p$ orbitals centered on the same atom:
\begin{equation}
\phi_m=\psi_{2p_x},
\qquad
\phi_n=\psi_{2p_y},
\qquad 
\phi_l=\psi_{1s}.
\label{eq:hydrogen-state-assignment}
\end{equation}
The normalized hydrogenic orbitals are
\begin{equation}
\begin{aligned}
\psi_{1s}(\mathbf r)
&=
\frac{1}{\sqrt{\pi}}e^{-r},\\
\psi_{2p_x}(\mathbf r)
&=
\frac{x}{4\sqrt{2\pi}}e^{-r/2},\\
\psi_{2p_y}(\mathbf r)
&=
\frac{y}{4\sqrt{2\pi}}e^{-r/2},
\end{aligned}
\label{eq:hydrogen-orbitals}
\end{equation}
where $r=|\mathbf r|$.

The velocity matrix element between the $1s$ and $2p_x$ states is
\begin{equation}
\begin{aligned}
\langle 1s|\hat v_x|2p_x\rangle
&=
-i\int
\psi_{1s}(\mathbf r)
\frac{\partial\psi_{2p_x}(\mathbf r)}{\partial x}
\,d^3r\\
&=
-\frac{i}{4\sqrt{2}\pi}
\int e^{-3r/2}
\left(
1-\frac{x^2}{2r}
\right)d^3r\\
&=
-i\frac{16\sqrt{2}}{81}.
\end{aligned}
\label{eq:hydrogen-vx-matrix-element}
\end{equation}
The remaining Cartesian components vanish by spatial symmetry.
Using the same procedure for the $1s$ and $2p_y$ states, the relevant velocity matrix elements are
\begin{equation}
\begin{aligned}
\mathbf v_{1s,2p_x}
&=
\langle 1s|\hat{\mathbf v}|2p_x\rangle
=
-i\frac{16\sqrt{2}}{81}\mathbf e_x,\\
\mathbf v_{2p_y,1s}
&=
\langle 2p_y|\hat{\mathbf v}|1s\rangle
=
+i\frac{16\sqrt{2}}{81}\mathbf e_y,
\end{aligned}
\label{eq:hydrogen-velocity-matrix-elements}
\end{equation}
where $\mathbf e_x$ and $\mathbf e_y$ are Cartesian unit vectors.

Following the intermediate-state decomposition in Eq.~\eqref{eq:vortex-intermediate-kernel}, the contribution of the $1s$ state to the current kernel between $2p_x$ and $2p_y$ is
\begin{equation}
\mathbf K_{2p_x,2p_y}^{1s}(\mathbf r)
=
\psi_{2p_y}(\mathbf r)
\psi_{1s}(\mathbf r)
\mathbf v_{1s,2p_x}
+
\psi_{2p_x}(\mathbf r)
\psi_{1s}(\mathbf r)
\mathbf v_{2p_y,1s}.
\label{eq:hydrogen-current-kernel-definition}
\end{equation}
Substituting the hydrogenic orbitals and velocity matrix elements gives
\begin{equation}
\mathbf K_{2p_x,2p_y}^{1s}(\mathbf r)
=
i\frac{4}{81\pi}
e^{-3r/2}
\left(
-y\mathbf e_x+x\mathbf e_y
\right).
\label{eq:hydrogen-px-py-kernel}
\end{equation}

\begin{figure}[!t]
\centering
\includegraphics[width=\linewidth]{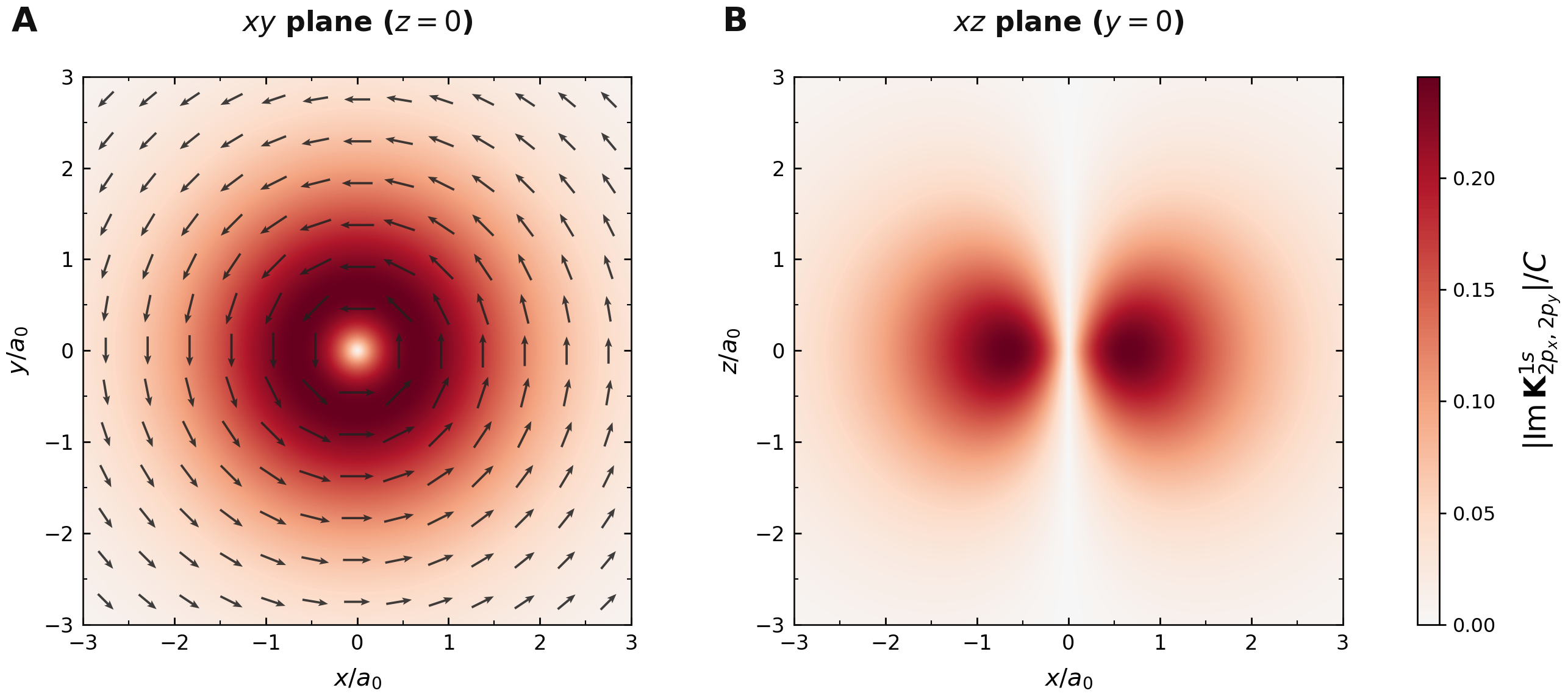}
\caption{
Spatial structure of the hydrogenic current kernel.
The normalized kernel is defined as $\widetilde{\mathbf K}=\operatorname{Im}\mathbf K/C$, with $C=4/(81\pi)$.
(A) Magnitude $|\widetilde{\mathbf K}|$ in the $xy$ plane at $z=0$, with arrows indicating its azimuthal direction.
(B) Corresponding magnitude in the $xz$ plane at $y=0$, showing the out-of-plane localization of the kernel circulating in the $xy$ plane.
Both panels share the same linear color scale.
Coordinates are given in units of the Bohr radius $a_0$.
}
\label{fig:hydrogenic-current-kernel}
\end{figure}

To examine the spatial structure of this kernel, we introduce cylindrical coordinates,
$x=R\cos\varphi$ and $y=R\sin\varphi$.
Since $-y\mathbf e_x+x\mathbf e_y=R\mathbf e_\varphi$, Eq.~\eqref{eq:hydrogen-px-py-kernel} becomes
\begin{equation}
\mathbf K_{2p_x,2p_y}^{1s}(R,\varphi,z)
=
i\frac{4}{81\pi}
R e^{-\frac{3}{2}\sqrt{R^2+z^2}}
\mathbf e_\varphi.
\label{eq:hydrogen-px-py-kernel-cylindrical}
\end{equation}
As shown in Fig.~\ref{fig:hydrogenic-current-kernel}, coupling between two orthogonal $p$ orbitals through a common $s$ state naturally produces an azimuthal current contribution.

\section{Additional Results}\label{sec:si-additional-results}

Unless otherwise stated, the QE calculations used to generate the electronic-structure data for CPA-NAMD-extF employ the PBE exchange-correlation functional and SG15 PBE norm-conserving pseudopotentials. A plane-wave cutoff of 50 Ry and $\Gamma$-point sampling are adopted throughout. Ground-state electronic structures, wave-function overlaps, and velocity matrix elements are generated with QE during BOMD.

\subsection{Benchmark of the Time-Dependent Dipole Response}\label{sec:si-dipole-benchmark}

We first benchmark the time-dependent dipole response against PWmat rt-TDDFT in order to validate the charge-density response.
Three systems are considered: an $\mathrm{H_2CO_3}$ molecule, a conjugated 1D chain $\mathrm{C}_{14}\mathrm{H}_{12}\mathrm{BN}$, and monolayer 2D $\mathrm{WS_2}$. 
The same ionic trajectories and external fields are used in both TDDFT and CPA-NAMD-extF calculations, thereby eliminating differences associated with ionic motion and enabling a direct comparison of the electronic dipole dynamics.

All systems are charge neutral and non-spin-polarized. 
CPA-NAMD-extF uses the same ionic trajectories as the PWmat rt-TDDFT calculations. 
The PWmat trajectories have a time step of 0.1 fs, from which one atomic configuration is sampled every 1 fs for the QE BOMD calculations, yielding 101 configurations over 0--100 fs. 
Spatial symmetry is disabled, and the nonlocal pseudopotential contribution is included in the velocity matrix elements. 
The system sizes, temperatures, and electronic-state windows are summarized in Table~S1.

\begin{table}[htbp]
\centering
\begin{tabular}{lcccc}
\toprule
System & No. of atoms & No. of valence electrons & Temperature (K) & Active window \\
\midrule
$\mathrm{C}_{14}\mathrm{H}_{12}\mathrm{BN}$ & 28 & 76 & 300 & 15--150 \\
$\mathrm{H_2CO_3}$ & 6 & 24 & 500 & 1--41 \\
$\mathrm{WS}_2$ & 48 & 416 & 500 & 97--322 \\
\bottomrule
\end{tabular}
\caption{Structural information and active electronic-state windows used in the dipole benchmark.}
\label{tab:dipole-benchmark-systems}
\end{table}

CPA-NAMD-extF uses an electronic time step of 0.1 fs, corresponding to 10 electronic steps between adjacent QE BOMD configurations. 
The total propagation time is 100 fs.
The external field is polarized along $z$ and has a Gaussian-envelope sinusoidal form,
\begin{equation}
\mathbf E(t)
=
E_0
\exp\left[-\frac{(t-t_c)^2}{\sigma^2}\right]
\sin(\omega_0 t+\phi)\,
\hat{\mathbf z},
\label{eq:dipole-benchmark-field}
\end{equation}
with $E_{\mathrm{ph}}=\hbar\omega_0$. Here, $\sigma$ denotes the Gaussian width parameter. The field parameters are listed in Table~S2. PWmat and CPA-NAMD-extF use the same field waveform and parameters.

\begin{table}[htbp]
\centering
\begin{tabular}{cccccccc}
\toprule
System & Field direction & $I_0$ (W cm$^{-2}$) & $E_0$ (V \AA$^{-1}$) & $t_c$ (fs) & $\sigma$ (fs) & $E_{\mathrm{ph}}$ (eV) & $\phi$ \\
\midrule
$\mathrm{C}_{14}\mathrm{H}_{12}\mathrm{BN}$ & $+z$ & $5.0\times10^{12}$ & 0.6138 & 50 & 15 & 2.00 & $\pi/2$ \\
$\mathrm{H_2CO_3}$ & $+z$ & $4.0\times10^{12}$ & 0.5490 & 50 & 15 & 2.00 & $\pi/2$ \\
$\mathrm{WS}_2$ & $+z$ & $5.0\times10^{12}$ & 0.6138 & 50 & 15 & 2.00 & $\pi/2$ \\
\bottomrule
\end{tabular}
\caption{External-field parameters used in the dipole benchmark.}
\label{tab:dipole-benchmark-field}
\end{table}

\begin{figure}[htbp]
\centering
\includegraphics[width=\linewidth]{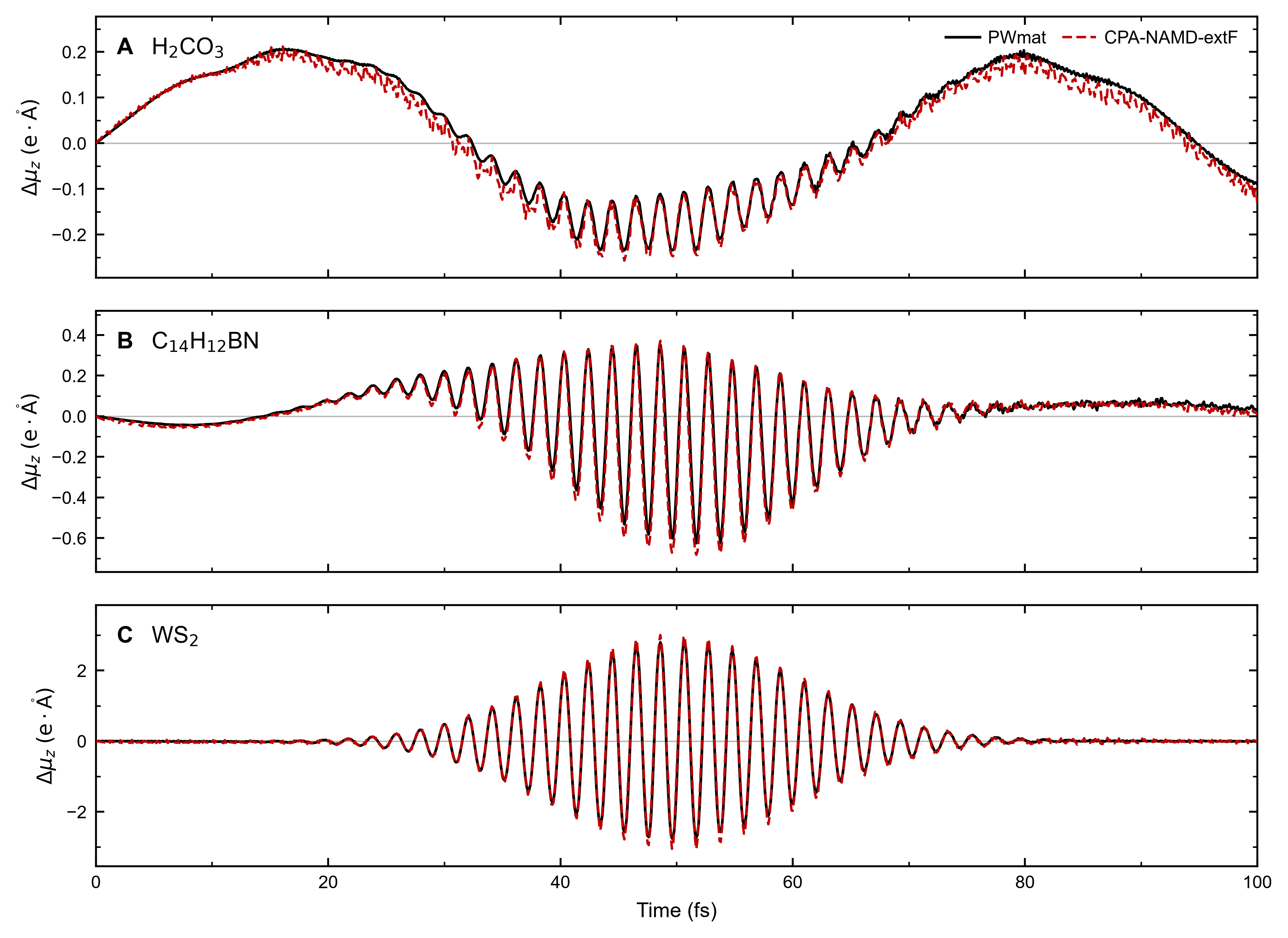}
\caption{Time-dependent total dipole change $\Delta\mu_z(t)$ obtained from PWmat rt-TDDFT and CPA-NAMD-extF. Panels A--C correspond to $\mathrm{H_2CO_3}$, $\mathrm{C}_{14}\mathrm{H}_{12}\mathrm{BN}$, and $\mathrm{WS_2}$, respectively. Black solid and red dashed curves denote the PWmat and CPA-NAMD-extF results, respectively.}
\label{fig:dipole-benchmark}
\end{figure}

For CPA-NAMD-extF, the corresponding vector potential is
\begin{equation}
\mathbf A(t)
=
-c\int_0^t\mathbf E(t')\,dt',
\qquad
\mathbf A(0)=0.
\label{eq:dipole-benchmark-vector-potential}
\end{equation}
Using the conventional sign definition of the electric dipole moment, the electronic, ionic, and total dipole moments are
\begin{equation}
\boldsymbol{\mu}_{\mathrm{el}}(t)
=
-\int \rho(\mathbf r,t)\mathbf r\,d^3r,
\qquad
\boldsymbol{\mu}_{\mathrm{ion}}(t)
=
\sum_{\kappa} Z_{\kappa}^{\mathrm{val}}\mathbf R_{\kappa}(t),
\qquad
\boldsymbol{\mu}_{\mathrm{tot}}(t)
=
\boldsymbol{\mu}_{\mathrm{el}}(t)+\boldsymbol{\mu}_{\mathrm{ion}}(t).
\label{eq:dipole-total}
\end{equation}
Because both methods use the same ionic trajectory, their ionic contributions are identical. We therefore compare the change in the total dipole moment, $\Delta\boldsymbol{\mu}(t)=\boldsymbol{\mu}_{\mathrm{tot}}(t)-\boldsymbol{\mu}_{\mathrm{tot}}(0)$.
As shown in Fig.~\ref{fig:dipole-benchmark}, these three calculations show very close agreement.
The dominant oscillation frequencies, phases, and temporal envelopes are well reproduced, with no systematic drift over 100 fs. The remaining discrepancies are primarily small local amplitude differences.

\subsection{Benchmark of the Real-Time Current Response}\label{sec:si-current-benchmark}

We next benchmark the real-time current against SALMON rt-TDDFT using a fixed $3\times3\times1$ monolayer $\mathrm{MoS_2}$ supercell. Both calculations use the same atomic structure, simulation cell, vacuum thickness, and external field. Fixing the ions isolates the electronic propagation and current evaluation.

\begin{figure}[htbp]
\centering
\includegraphics[width=0.49\linewidth]{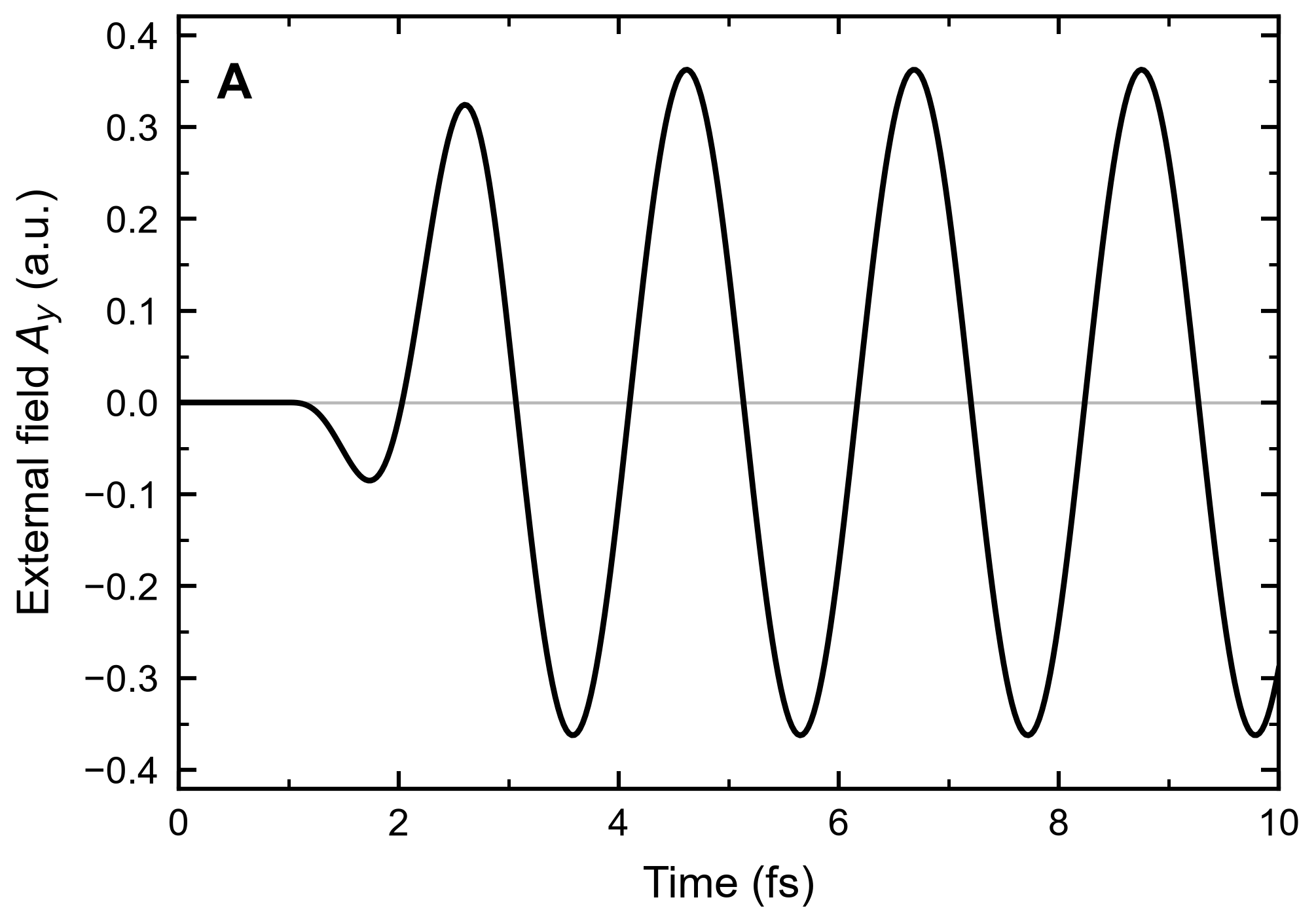}\hfill
\includegraphics[width=0.49\linewidth]{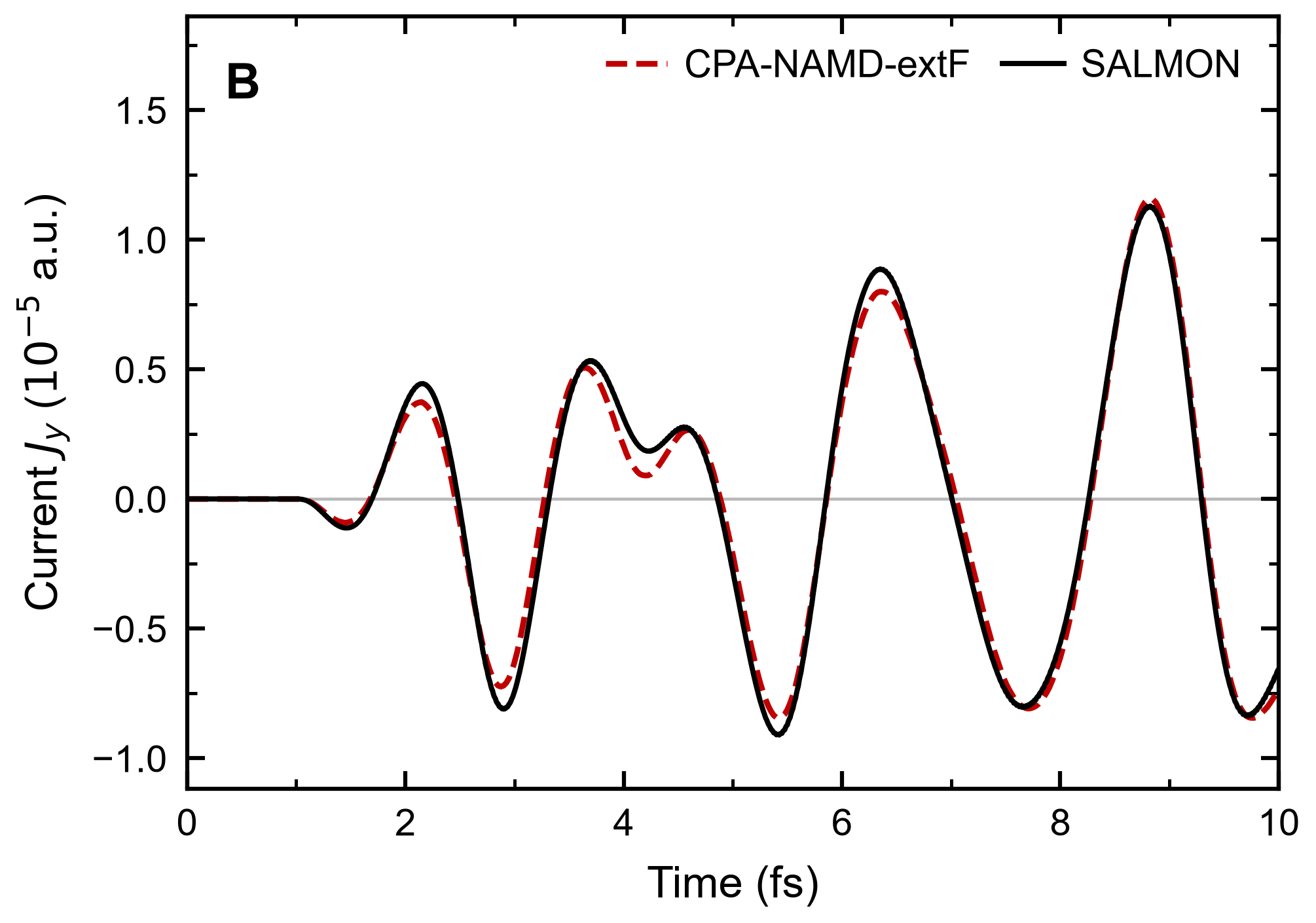}
\caption{Finite-field current benchmark for monolayer $\mathrm{MoS_2}$. (A) Applied vector potential $A_y(t)$. (B) Current $J_y(t)$ from CPA-NAMD-extF and SALMON RT-TDDFT. 
Black solid and red dashed curves denote SALMON and CPA-NAMD-extF, respectively.
The current is plotted in units of $10^{-5}$ a.u.}
\label{fig:mos2-current-benchmark}
\end{figure}

A monochromatic field polarized along $y$ is applied using Eq.~\eqref{eq:monochromatic-vector-potential}. The parameters are $E_0=0.01~\mathrm{V}\,\text{\AA}^{-1}$, $E_{\mathrm{ph}}=2.0~\mathrm{eV}$, $\phi=\pi/2$, $t_{\mathrm{on}}=1~\mathrm{fs}$, and a 2 fs switching ramp. The applied vector potential $A_y(t)$ is shown in Fig.~\ref{fig:mos2-current-benchmark}(A).

The CPA-NAMD-extF active window contains KS states 1--1068. The QE BOMD time step is 0.5 fs and the electronic time step is 0.025 fs, giving 20 electronic steps per BOMD step. The ionic coordinates are identical at all reference time points. SALMON uses an electronic time step of 0.2 as. The CPA-NAMD-extF current is evaluated from Eq.~\eqref{eq:current}, including the paramagnetic, diamagnetic, and finite-band adiabatic-correction terms. We compare the $y$ component, $J_y(t)$.

The two currents show nearly identical oscillation frequencies, phases, and peak positions, with only small local amplitude differences. 
These results support the accuracy of the CPA-NAMD-extF real-time current evaluation for this periodic benchmark.

\subsection{Benchmark of the Linear Optical Response}\label{sec:si-linear-optical-benchmark}

Finally, we benchmark the linear optical response of crystalline Si against PWmat RT-TDDFT. A $4\times4\times4$ supercell constructed from the two-atom primitive cell is used, containing 128 Si atoms.
A weak $\delta$-kick is applied along $[111]$. The kick magnitude is $0.002~\mathrm{bohr}^{-1}$,
\begin{equation}
\Delta\mathbf{k}
=
0.002\frac{(1,1,1)}{\sqrt{3}}
~\mathrm{bohr}^{-1},
\label{eq:si-delta-kick}
\end{equation}
corresponding to $\Delta k_x=\Delta k_y=\Delta k_z=1.155\times10^{-3}~\mathrm{bohr}^{-1}$. The kick is applied at $t_0=0.1~\mathrm{fs}$.

PWmat first propagates for 0.2 fs with a 1 as time step, followed by 10 fs with a 0.01 fs time step. 
For CPA-NAMD-extF, the fixed-ion QE reference data are generated at 0.05 fs intervals. 
The total propagation time is 15 fs.

The time-dependent current is processed using the linear-response procedure described above. The CPA-NAMD-extF spectra are linearly interpolated onto the PWmat energy grid, and the comparison is performed over the common energy range of 0.4--20 eV.

\begin{figure}[htbp]
\centering
\includegraphics[width=0.49\linewidth]{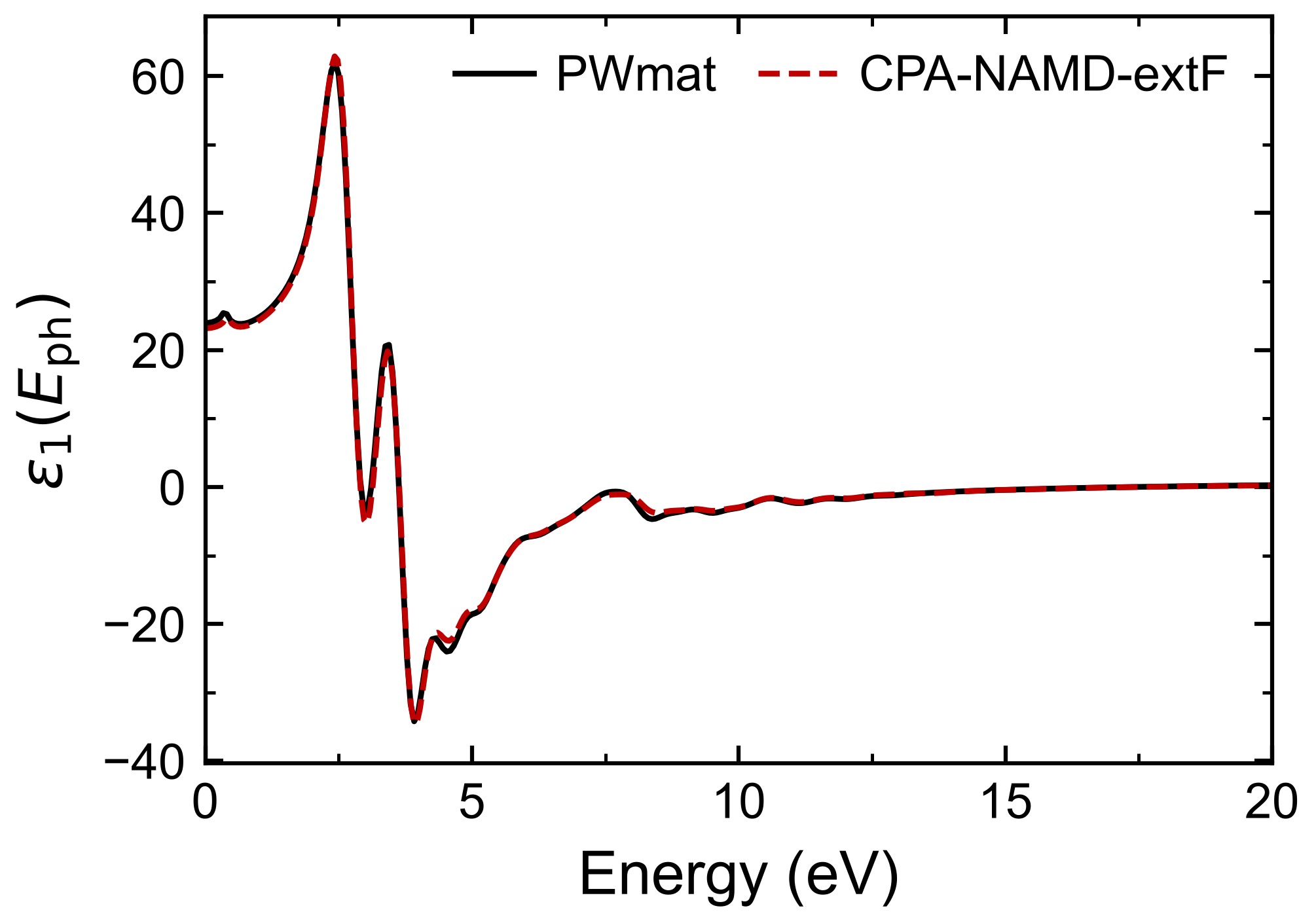}\hfill
\includegraphics[width=0.49\linewidth]{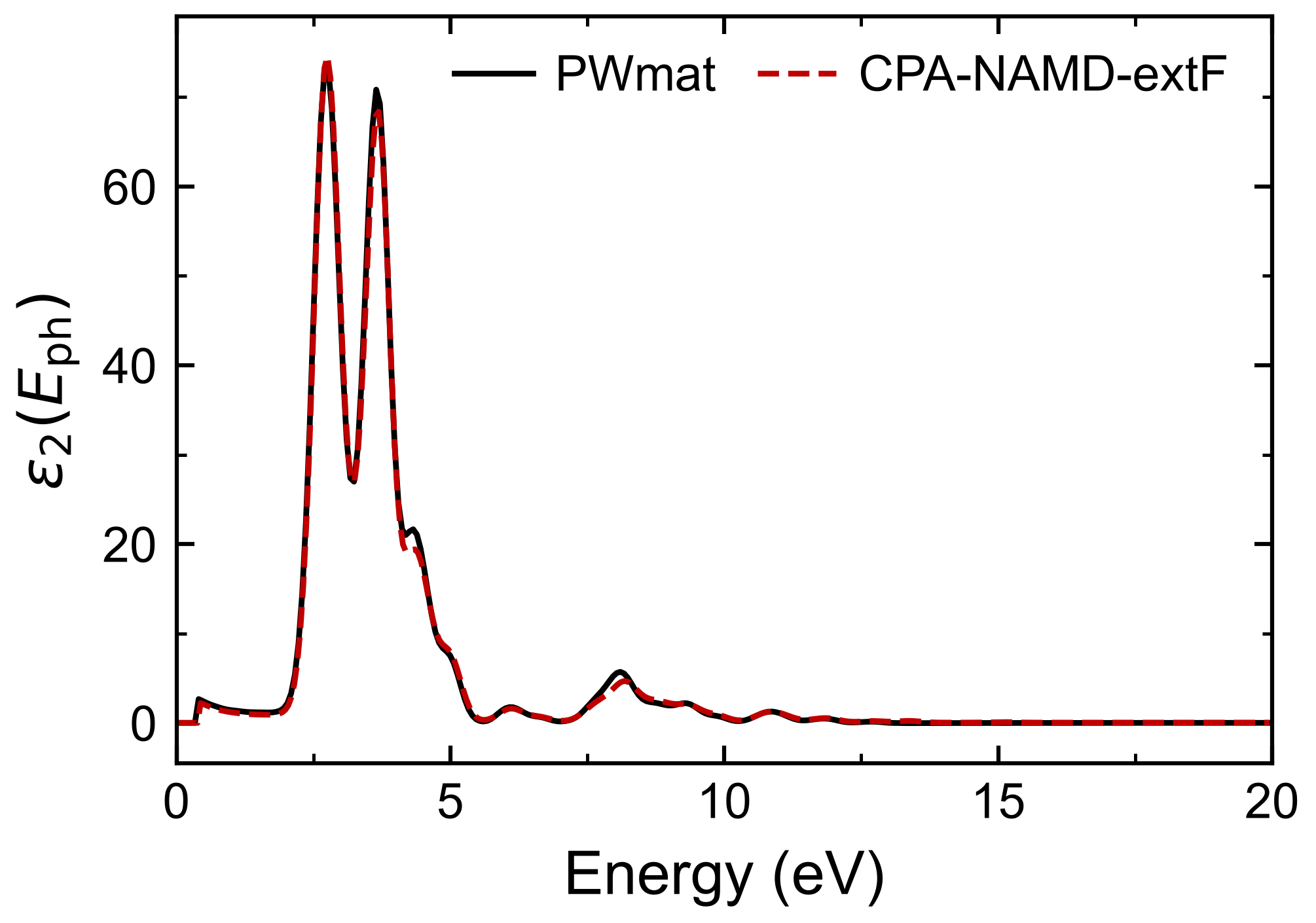}
\caption{Linear optical-response benchmark for crystalline Si. The left and right panels show $\varepsilon_1(E_{\mathrm{ph}})$ and $\varepsilon_2(E_{\mathrm{ph}})$ obtained from PWmat rt-TDDFT and CPA-NAMD-extF. Black solid and red dashed curves denote PWmat and CPA-NAMD-extF, respectively.}
\label{fig:si-linear-benchmark}
\end{figure}

These two calculations show negligible differences as shown in Fig.~\ref{fig:si-linear-benchmark}.
Some very small differences are mainly confined to the relative intensities of weak high-energy features. 
These results support the reliability of CPA-NAMD-extF for describing the linear optical response of crystalline Si under the present benchmark conditions.

\subsection{Weak-Field Scaling of the Second-Order Response}\label{sec:si-weak-field-convergence}

To verify the quadratic scaling of the DC current with field amplitude, we perform calculations using different field strengths but with the same thermal configurations of the 300 and 450 K trajectories.
In the weak-field regime, the even-order current satisfies
\begin{equation}
J_{\mathrm{even},z}(t;E_0)
=
E_0^2J_z^{[2]}(t)
+
\mathcal O(E_0^4).
\label{eq:weak-field-current}
\end{equation}
Consequently, the time-averaged DC current satisfies
\begin{equation}
J_{\mathrm{DC},z}(E_0)
=
\sigma_{zzz}^{(2)}E_0^2
+
\mathcal O(E_0^4).
\label{eq:weak-field-dc-current}
\end{equation}
Therefore, the DC current should scale linearly with $E_0^2$ within the weak-field regime.
For each temperature, we select 13 starting configurations from the NVE trajectory between $1.04$ and $2.00~\mathrm{ps}$ at intervals of $80~\mathrm{fs}$, giving a total of 26 starting configurations.
For each configuration, calculations with positive and negative field amplitudes, together with a zero-field reference, are performed for $E_0=0.005$, $0.010$, and $0.020~\mathrm{V}\,\text{\AA}^{-1}$.
All other computational parameters, including the photon energy of $4.0~\mathrm{eV}$, are kept unchanged.
The even-order current is averaged over 16 complete optical cycles to obtain $J_{\mathrm{DC},z}$ for each configuration.
The resulting DC currents are then averaged over the 13 configurations to obtain $\bar{J}_{\mathrm{DC},z}$ at each field amplitude.
This procedure yields 39 DC-current values at each temperature (13 configurations and 3 field amplitudes).

\begin{figure}[!t]
\centering
\includegraphics[width=\linewidth]
{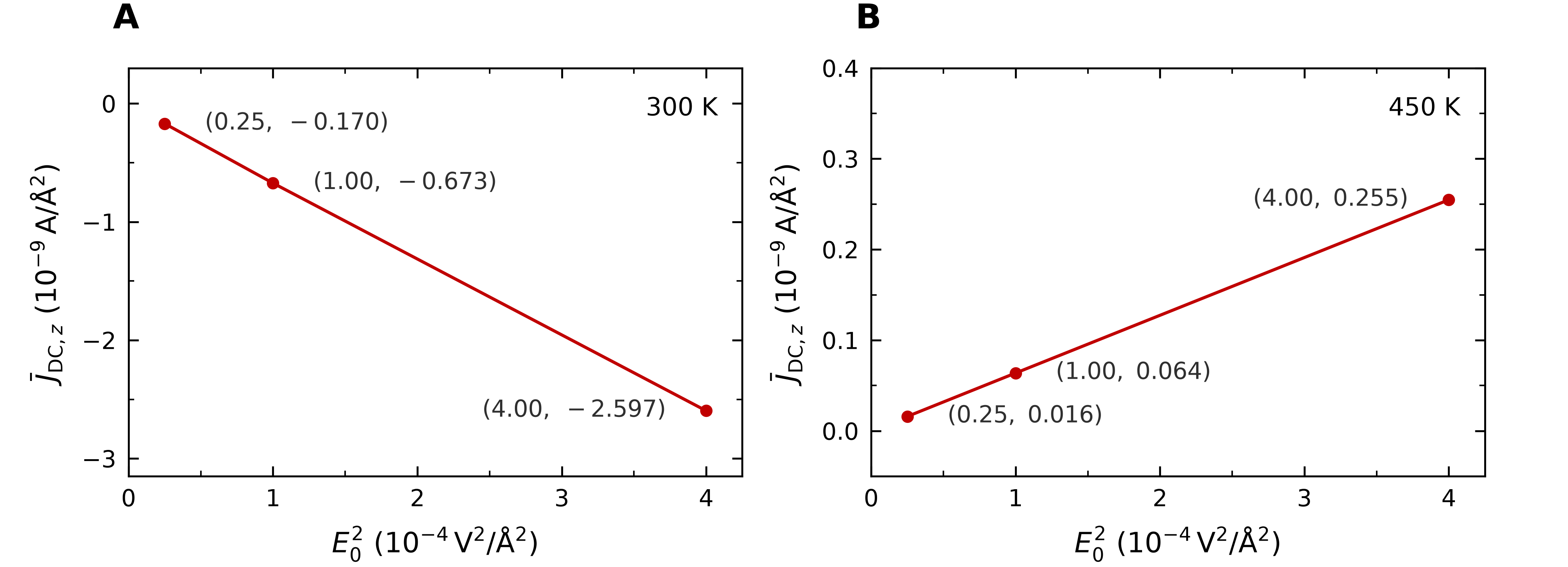}
\caption{
Linear scaling of the configuration-averaged DC current with $E_0^2$.
(A) Results for the 300 K NVE trajectory.
(B) Results for the 450 K NVE trajectory.
For each temperature, the same 13 thermal configurations are used at
$E_0=0.005$, $0.010$, and $0.020~\mathrm{V}\,\text{\AA}^{-1}$.
Red filled circles represent the configuration-averaged DC currents at each field amplitude.
The vertical axis shows $\bar{J}_{\mathrm{DC},z}$.
All calculations use a photon energy of $4.0~\mathrm{eV}$.
The even-order current is averaged over 16 complete optical cycles, corresponding to the sampled time window from $3.1$ to $19.6~\mathrm{fs}$.
The production calculations reported in the main text use $E_0=0.010~\mathrm{V}\,\text{\AA}^{-1}$.
The lines are guides to the eye.
}
\label{fig:weak-field-convergence}
\end{figure}

As shown in Fig.~\ref{fig:weak-field-convergence}, the configuration-averaged DC current $\bar{J}_{\mathrm{DC},z}$ exhibits approximately linear scaling with $E_0^2$ at both temperatures.
At 300 K, the configuration-averaged DC current remains negative, whereas at 450 K, it is positive with a much smaller magnitude.
The approximately linear scaling indicates that the even-order current is dominated by the second-order contribution in the weak-field regime.
These results also support the use of $E_0=0.010~\mathrm{V}\,\text{\AA}^{-1}$ for the second-order response calculations reported in the main text.

\clearpage
\bibliography{shared/references}